\documentclass[11pt,a4paper]{article}

\usepackage{jheppub}
\usepackage{pdfsync}

\usepackage{epsfig,epsf}
\usepackage{amsmath}
\usepackage{amsthm}
\usepackage{amsfonts}
\usepackage{amssymb}
\usepackage{dsfont}
\usepackage{marvosym}
\usepackage{multirow}

\RequirePackage{graphicx}
\RequirePackage{flushend}
\RequirePackage[numbers,sort&compress]{natbib}

\usepackage{slashed}
\usepackage{epstopdf}
\usepackage[active]{srcltx}

\def\II{\hbox{{1}\kern-.25em\hbox{l}}}

\newcounter{VBQ}

\title{Two-photon processes in conformal QCD:
Resummation of the descendants of leading-twist operators}

\author[a]{V. M. Braun,}
\author[b]{Yao Ji,}
\author[c,d,a]{and A. N.  Manashov\,}

\subheader{
\begin{flushright}
{\small
DESY 20-189 \\ SI-HEP-2020-26
}
\end{flushright}
}

\affiliation[a]{
   Institut f\"ur Theoretische Physik, Universit\"at
   Regensburg,  D-93040 Regensburg, Germany}

\affiliation[b]{Theoretische Physik 1, Naturwissenschaftlich-Technische Fakult{\"a}t,
Universit{\"a}t Siegen, 57068 Siegen,\\ Germany}

\affiliation[c]{
   Institut f\"ur Theoretische Physik, Universit\"at Hamburg,
   D-22761 Hamburg, Germany}

\affiliation[d]{ St.Petersburg Department of Steklov
Mathematical Institute,
191023 St.Petersburg, Russia
 }

\emailAdd{vladimir.braun@physik.ur.de}

\emailAdd{yao.ji@uni-siegen.de}

\emailAdd{alexander.manashov@desy.de}

\abstract{
Using some techniques of conformal field theories, we find a closed expression for the contribution
of leading twist operators and their descendants, obtained by adding total derivatives, to the
operator product expansion (OPE) of
two electromagnetic currents in QCD.
Our expression resums contributions of all twists and to all orders in perturbation theory
up to corrections proportional to the QCD $\beta$-function.
At tree level and to twist-four accuracy, our result
agrees with the expression derived earlier by a different method.
The results are directly applicable to deeply-virtual Compton scattering and, e.g.,
$\gamma\gamma^\ast$ annihilation in two mesons.
As a byproduct, we derive a simple representation for the OPE of two  scalar currents
that is convenient for applications.
       }


\begin{document}

\maketitle

\newpage

\section{Introduction}
Hard exclusive processes with one or two virtual photons are attracting increasing attention because of quality of the experimental
data that are already arriving and expected within a few years from Jlab 12 GeV upgrade~\cite{Dudek:2012vr}, SuperKEKB
\cite{Kou:2018nap}, and later from the EIC~\cite{Accardi:2012qut}. The prime motivation for the present study is provided by the
deeply-virtual Compton scattering (DVCS), but the results are also relevant for reactions of the type $\gamma\gamma^\ast \to
\pi\pi$ etc.

DVCS is one of the main processes for spatial imaging of partons inside the nucleon. As the spatial position of partons is Fourier
conjugate to the momentum transfer to the nucleon in the scattering process, the resolving power of DVCS is directly limited by the
range of the invariant momentum transfer $t$ that can be used in the analysis.  Since factorization for DVCS includes power
corrections in $t/Q^2$, theoretical control over these corrections is of paramount importance.
At this time, a calculation of kinematic power corrections to DVCS to the twist-four accuracy, $\sim t/Q^2$ and ${\sim m^2/Q^2}$,
is available~\cite{Braun:2014sta} following the approach developed in~\cite{Braun:2011zr,Braun:2011dg,Braun:2012bg,Braun:2012hq}. A
typical size of kinematic corrections is of order 10\% for asymmetries, but they can be as large as 100\% for the total cross
section in certain kinematics. These corrections can significantly impact the extraction of GPDs from the data and have to be taken
into account~\cite{Defurne:2015kxq,Defurne:2017paw}. In this paper we develop an approach that allows to resum the corrections
$\sim (\sqrt{-t}/Q)^k$ and ${\sim (m/Q)^k}$ to all powers. Such all-order results are especially important for the newly emerging
subject of coherent DVCS from light nuclei \cite{Hattawy:2017woc}, in which case one needs to prove that QCD factorization is not
spoiled by the nucleus mass corrections, terms $\sim m_A/Q$.

From the theory point of view, the necessity to include kinematic power corrections is due to the well-known deficiency of the
``standard'' leading twist approximation: Violation of electromagnetic Ward identities and dependence of the results on the choice
of the  reference frame in the definition of the skewness parameter $\xi$ and the Compton form factors. A detailed discussion can
be found in~\cite{Braun:2014sta}. In both cases the corrections that restore the invariance are formally of subleading power in the
hard scale. They can be called ``kinematic'' as they do not involve new nonperturbative input, and have to be distinguished from
the ``genuine'' higher-twist corrections that arise from parton distributions of higher twist. Such higher-twist distributions are
very interesting by themselves as they carry unique information about parton correlations inside a hadron, but can only be accessed
after the kinematic effects are subtracted.

On a more formal level, the task can be formulated as follows. Let $\mathcal O^{\mu_1\ldots \mu_N}$ be local twist-two
operators. Matrix elements of these operators define moments of generalized parton distributions and the collinear factorization in
DVCS corresponds to taking into account contributions of all such operators with arbitrary spin $N$. The  ``kinematic''
approximation we are considering here is tantamount to taking into account
 contributions of {\it higher-twist} descendants of the twist-two operator $\mathcal O^{\mu_1\mu_2\ldots \mu_N}$,
\begin{align}
   \partial_{\mu_1}\mathcal O^{\mu_1\mu_2\ldots \mu_N}, \qquad
    \partial_{\mu_1}\partial_{\mu_2}\mathcal O^{\mu_1\mu_2\mu_3\ldots \mu_N}, \qquad
\partial^2 \mathcal O^{\mu_1 \ldots \mu_N}, \quad \text{etc.},
\label{intro:1}
\end{align}
where $\partial_\mu$ is a total derivative.  Matrix elements of these operators over states with equal momenta vanish; their
contributions are thus specific and endemic in reactions involving a momentum transfer between the initial  and final state
hadrons, DVCS being a prime example. It is necessary to calculate the coefficients with which these
operators enter the OPE of two electromagnetic currents. The usual method of calculating OPE coefficient functions ---  to
evaluate  both sides of the OPE on free quarks --- does not work here
since the matrix elements of the first two operators in \eqref{intro:1} (and similar ones with more derivatives) vanish for
on-mass-shell partons.

Consider, as an example, the spin-two quark-antiquark operator:
$
  O_{\mu\nu} = \frac12[\bar q \gamma_\mu\!
  \stackrel{\leftrightarrow}{D}_\nu \!q
  + (\mu\leftrightarrow\nu)]
$, which is nothing but  the quark contribution to the QCD energy-momentum tensor. Using QCD equations of motion (EOM) it is easy
to show that (for massless quarks)
\begin{align}
  \partial^\mu O_{\mu\nu} = 2ig \,\bar q  F_{\nu\mu}\gamma^\mu q\,,
\label{eq:puzzle}
\end{align}
where $g$ is the QCD coupling and $F_{\mu\nu}$ is the gluon field strength tensor.
The r.h.s. of this operator identity
is of order $g$; hence the matrix element of the l.h.s. over on-shell quarks vanishes at
leading order.
Eq.~\eqref{eq:puzzle} implies that the off-forward nucleon matrix element of the quark-gluon operator on the r.h.s. of this relation
is related to the matrix element of the leading twist operator
\begin{align}
2ig \,\langle p'|\bar q  F_{\nu\mu}\gamma^\mu q|p\rangle =
\langle p'|\partial^\mu \mathcal O_{\mu\nu}|p\rangle=i(p'-p)^\mu \langle p'| \mathcal O_{\mu\nu}|p\rangle
\end{align}
which defines the corresponding form-factor of the energy-momentum tensor \cite{Polyakov:2018zvc}, or, alternatively, the second
moment of the GPD $H(x,\xi)$, see, e.g., \cite{Tanaka:2018wea}. Thus, on the one hand, the corresponding contribution to DVCS is
expressed entirely in terms of twist-two GPDs (and is necessary to restore the Ward identities to twist-four accuracy). It is
therefore naturally interpreted as a part of the ``kinematic'' power correction. On the other hand, the matrix element of the
twist-four operator on the r.h.s. of Eq.~\eqref{eq:puzzle} involves a gluon field and is, naively, a measure of quark-gluon
correlations. It would be then tempting to attribute its contribution to the Compton tensor to the ``dynamical'' power correction.
This example shows that the separation between  ``kinematic'' power corrections from ``genuine'' quark-gluon contributions is
rather subtle.

\begin{figure}[t]
\centerline{\includegraphics[width=0.50\textwidth]{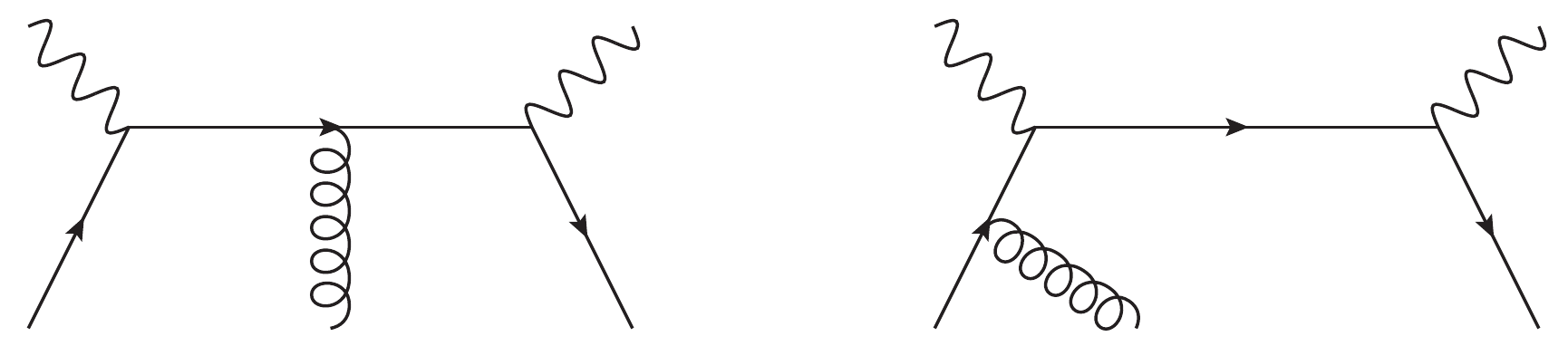}}
\caption{A quark-antiquark-gluon matrix element of the product of electromagnetic currents.
}
\label{fig:1}
\end{figure}

The corresponding technique was developed in \cite{Braun:2011zr,Braun:2011dg} and is based on
considering quark-antiquark gluon matrix elements such as shown in Fig.\,\ref{fig:1}, for which the separation
of the descendants of leading twist operators (``kinematic'' contributions)
from ``genuine'' quark-gluon contributions can be defined unambiguously.
In this way the OPE of two electromagnetic currents was derived in \cite{Braun:2011zr,Braun:2011dg}
in the ``kinematic'' approximation to twist-four accuracy.
%
The calculation of finite-$t$ and target mass corrections in DVCS  based  on this approach~\cite{Braun:2012hq, Braun:2014sta}
showed that they are quite sizable at low $Q^2$ and therefore taking into account contributions of all twists is quite desirable.
Unfortunately, the approach of \cite{Braun:2011zr,Braun:2011dg} becomes rather complicated beyond twist four approximation.

In this paper we calculate the kinematical power corrections of all twists using methods of  conformal field theories (CFTs). It is
well known that conformal symmetry allows one to fix the coefficients with which the descendant operators enter the
OPE~\cite{Ferrara:1971vh,Ferrara:1971zy,Ferrara:1973yt}. Thus in a  conformal field theory such coefficient functions  are related
to the coefficient functions of primary operators and can be obtained by considering the forward matrix elements. For QCD, this
would mean that kinematic corrections to DVCS amplitudes are unambiguously determined by the DIS coefficient functions.

Of course, QCD is not a conformal theory.
However, one can consider a modified theory, QCD in non-integer, $d=4-2\epsilon$, space-time dimensions and
fine-tune the strong coupling $\alpha_s$ for given $\epsilon$ to nullify the $\beta$-function
(Wilson-Fisher fixed point~\cite{Wilson:1973jj}) and make the theory  scale invariant.
 It is possible to show that conformal invariance  of correlation functions of gauge-invariant operators in this theory
is also restored~\cite{Braun:2018mxm}.
Observables calculated in the four-dimensional and critical QCD differ beyond leading order by terms
proportional to the QCD $\beta$ function, which, as a rule, appear starting from NNLO only.
They can be calculated, at least in principle \cite{Braun:2020yib}, but for higher twists are probably beyond the
accuracy of possible applications to GPD phenomenology so that we leave this question for further study.
At the tree level, there are no {differences between QCD in $d=4-2\epsilon$ and $d=4$ dimensions.}

As the main result, in this paper we construct the OPE for the product of two conserved vector currents in a generic CFT. The
similar expansion for the product of two scalar currents was derived long ago in Ref.~\cite{Ferrara:1971vh} whereas the
generalization to vector currents, to the best of our knowledge, has not been worked out. We find a simple representation for the
coefficient functions that is more explicit compared to~\cite{Ferrara:1971vh} and better suited for applications to high-energy
QCD. The contribution of a given twist-two operator and all its descendants of higher twist to
the OPE
{of two} vector currents is completely determined by symmetries up to the two normalization factors that, for QCD, can be related
to well-known coefficient functions $\mathbf{C}_2$ and $\mathbf{C}_L$ in DIS. At leading order, we reproduce in this way the
twist-four expression obtained in~\cite{Braun:2011zr}. This agreement is nontrivial as the two techniques are very different. It is
also important as a confirmation for the analysis in~\cite{Braun:2014sta}.

The presentation is organized as follows. In Sect.\ref{sect:triangles} we explain the main idea and introduce the shadow operator
formalism on the example of the OPE for the product of scalar currents. The result is presented in a form convenient for
applications. Details of this calculation can be found in App.~\ref{App:p-triangle}. Sect.~\ref{sect:vector} contains our main
result, the OPE for {two vector} currents in a generic conformal theory. In this Sect. we also calculate the remaining two
normalization coefficients in terms of the coefficient functions in DIS (in critical QCD). The relation of our result to a somewhat
different form of the conformal OPE at leading twist~\cite{Braun:2020yib} is established in App.~\ref{App:LT}. For readers'
convenience we collect main notations in App.~\ref{App:notations}.

\section{OPE of the product of scalar currents}\label{sect:triangles}

The general statement that the coefficient functions (CFs) of all descendants of primary operators in the OPE in a conformal theory
are fixed by the symmetry alone is known since long ago, and the corresponding expression for the product of two scalar operators
was first obtained in  Ref.~\cite{Ferrara:1971vh}.
In this section we re-derive  this result using a different approach based on the shadow operator formalism~\cite{Ferrara:1972uq}.
Our representation for the coefficient function is more explicit as compared to~\cite{Ferrara:1971vh} and seems to be
more convenient for applications. Throughout this section we will assume the Euclidean metric.

Let $\mathcal O_{\Delta_1}(x) \equiv \mathcal O_{1}(x) $ and
$\mathcal O_{\Delta_2}(x) \equiv \mathcal O_{2}(x) $ be scalar primary operators with scaling dimensions
$\Delta_1$ and $\Delta_2$. It means that these operators transform under dilatations
and inversion transformations $x'_\mu = x_\mu/x^2$  as
\begin{align}
\mathcal O_k(x)\mapsto \lambda^{\Delta_k} \mathcal O_k(\lambda x )\, , &&
\mathcal O_k(x)\mapsto (x'^2)^{\Delta_k} \mathcal O_k( x')\,,
\end{align}
respectively.
The OPE for the product of $\mathcal O_1$ and $\mathcal O_2$ runs over conformal (primary) symmetric traceless operators,
$\mathcal O^{\mu_1\ldots\mu_N}_{\Delta_N}$, where $\Delta_N$ is the corresponding scaling dimension, and their descendants, obtained
by adding total derivatives.
The primary operators $\mathcal O^{\mu_1\ldots\mu_N}_{\Delta_N}$ transform under inversion as follows
\begin{align}\label{tensor-law}
\mathcal O^{\mu_1\ldots\mu_N}_{\Delta_N}(x)\mapsto (x')^{{2\Delta_N}}\eta^{\mu_1}_{\ \nu_1}(x')\ldots
\eta^{\mu_N}_{\ \nu_N}(x')\mathcal O^{\nu_1\ldots\nu_N}_{\Delta_N}(x^\prime)\, ,
\end{align}
where the  tensor $\eta^{\mu\nu}$ has the form~\cite{Ferrara:1973yt}
\begin{align}
\eta^{\mu\nu}(x)=g^{\mu\nu}-
{2 x^\mu x^\nu}/{x^2}\,.
\end{align}
The OPE can be written, schematically, in the following form
\begin{align}\label{scalar:OPE}
\mathcal O_{1}(x_1)\mathcal O_{2}(x_2)=\sum_{N,\Delta_N} C_{\mu_1\ldots\mu_N}^{\Delta_N}(x_{12},\partial)\,
    \mathcal O^{\mu_1\ldots\mu_N}_{\Delta_N}(x)\, ,
\end{align}
where { $$x_{12}=x_1-x_2\, , \qquad x=(x_1+x_2)/2\, ,$$ and $\partial_\mu\equiv \partial/\partial x^\mu$ for later use}. The
coefficient function $C_{\mu_1\ldots\mu_N}^{\Delta_N}(x_{12},\partial)$ is a series in powers of the derivative and effectively
sums up contributions of all descendants of the form $\mathcal O^{\mu_1\ldots\mu_N}_{\Delta_N}$, cf. Eq.~\eqref{intro:1}
\begin{align}
C^{\mu_1\ldots\mu_N}_{\Delta_N}(x_{12},\partial)=C_{N,\Delta_N}
\frac{x_{12}^{\mu_1}\ldots x_{12}^{\mu_N}}{|x_{12}|^{\Delta_1+\Delta_2-\Delta_N+N}} +\mathcal O(\partial)\,.
\end{align}
One can always assume that the coefficient function is a traceless tensor in Lorentz indices
since the trace terms vanish after contraction with the operator
$\mathcal O^{\mu_1\ldots\mu_N}_{\Delta_N}$.

The main statement is that the functional form of the coefficient function  $C_{\mu_1\ldots\mu_N}^{\Delta_N}(x_{12},\partial)$
(i.e. including all powers of the derivative) is uniquely fixed by the transformation properties of the
operators on both sides of Eq.~\eqref{scalar:OPE}, up to an overall normalization constant~\footnote{
It is expected that the normalization constant only depends on a few parameters which specify the critical
point, see~\cite{Poland:2018epd} for a review.}. %
Explicit expression was first obtained in Ref.~\cite{Ferrara:1971vh} using the six-dimensional embedding formalism
(for the further development of this technique see \cite{Costa:2011mg}).
Here we use a different approach based on the shadow operator formalism~\cite{Ferrara:1972uq}.

The starting point is to consider the correlation function on both sides of the OPE,  Eq.~\eqref{scalar:OPE},
with a particular primary operator $\mathcal O^{\mu_1\ldots\mu_N}_{\Delta_N}(x_3)$.
It is well known~\cite{Polyakov:1970xd} that the two-point correlation function of two primary operators
with different scaling dimensions vanishes. Thus only one term in the sum survives, and one obtains
\begin{align}\label{first:triangle}
\left\langle\mathcal O_{1}(x_1)\mathcal O_{2}(x_2)\mathcal O^{\mu_1\ldots\mu_N}_{\Delta_N}(x_3)\right\rangle =
C^{\nu_1\ldots\nu_N}_{\Delta_N}(x_{12},\partial)\,
    \left\langle
        \mathcal O^{\nu_1\ldots\nu_N}_{\Delta_N}(x) \,
            \mathcal O^{\mu_1\ldots\mu_N}_{\Delta_N}(x_3)
     \right\rangle.
\end{align}
The two-point correlation function on the r.h.s. and the three-point correlation function  on the l.h.s.
of this relation are both fixed by conformal invariance up to a normalization constant.
The two-point function has the form
\begin{align}
\label{2point}
\left\langle
         \mathcal O^{\vec{\mu}_N}_{\Delta_N}(x)\mathcal O^{\vec{\nu}_N}_{\Delta_N}(x_3)\right\rangle
=  c_N \mathcal{D}_{\Delta_N}^{\vec{\mu}_N\,\vec{\nu}_N}(x-x_3),
\end{align}
where $\vec{\mu}_N=(\mu_1,\ldots,\mu_N)$, etc.,  and
\begin{align}
\mathcal{D}_{\Delta_N}^{\vec{\mu}_N\,\vec{\nu}_N}(x-y)=\frac1{|x-y|^{2\Delta_N}}\Biggl(
\frac1{N!} \sum_{\sigma\in S_N}{\prod_{i=1}^N \eta^{\mu_{\sigma(i)}\nu_i}(x-y)
-\text{traces}}
\Biggr),
\label{Dfunction}
\end{align}
where the sum is taken over all permutations of $N$ elements $(1,\ldots,N)$. The normalization constant $c_N$ is not fixed by the
symmetry and depends on a theory.

The three-point function can be written as
\begin{align}\label{3point}
\left\langle \mathcal O_1(x_1) \mathcal O_2(x_2) \mathcal O_{\Delta_N}^{\vec{\mu}_N}(x_3) \right\rangle
&= c'_N T^{\vec{\mu}_N}_{\Delta_N}(x_1,x_2,x_3),
\end{align}
where
\begin{align}
\label{triangle1}
T^{\vec{\mu}_N}_{\Delta_N}(x_1,x_2,x_3)
&
=\frac{
\Lambda^{\vec{\mu}_N}(x_1,x_2,x_3)
}{|x_{12}|^{\Delta_1+\Delta_2-\Delta_N+N}
|x_{13}|^{\Delta_1+\Delta_N-N-\Delta_2} |x_{23}|^{\Delta_2+\Delta_N-N-\Delta_1}}\,,
\end{align}
with
\begin{align}\label{Lambda:def}
\Lambda^{\vec{\mu}_N}(x_1,x_2,x_3)=\prod_{k=1}^N\Lambda^{\mu_k}(x_1,x_2,x_3)-\text{traces}, &&
\Lambda^\mu(x_1,x_2,x_3)=\frac{x_{13}^\mu}{x_{13}^2}-\frac{x_{23}^\mu}{x_{23}^2}\,,
\end{align}
and another normalization constant, $c'_N$.

The convolution of two  $\mathcal D$-functions defined in Eq.~\eqref{Dfunction} with the dimensions $\Delta_N$ and
 $\widetilde \Delta_N = d-\Delta_N$ (the shadow dimension) gives~\cite{Fradkin:1978pp}
\begin{align}\label{twoDD}
\int d^dy\,
\mathcal{D}_{\Delta_N}^{\vec{\mu}_N\,\vec{\nu}_N}(x_1-y)
\mathcal{D}_{\widetilde \Delta_N}^{\vec{\nu}_N\,\vec{\rho}_N}(y-x_2)
= D_N(\Delta_N)\, \delta^{(d)}(x_1-x_2)\, I^{\vec{\mu}_N\, \vec{\rho}_N}\, ,
\end{align}
where $I^{\vec{\mu}_N\, \vec{\rho}_N}$ is the traceless and symmetric tensor in both sets of indices,
\begin{align}
I^{\vec{\mu}_N\, \vec{\rho}_N}=\frac1{N!}\Biggl(\sum_{\sigma\in S_N} \prod_{k=1}^N g^{\mu_{\sigma(k)},\rho_k}
-\text{traces}
\Biggr),
\end{align}
{where $\sigma$ denotes a permutation of $N$ elements $\{1,2,\ldots, N\}$,} and
\begin{align}
\label{D_N}
D_N(\Delta_N) &=
\pi^{d}
\frac{\Gamma(d/2-\Delta_N)}{\Gamma(\Delta_N-1)(\Delta_N+N-1)}\cdot
\frac{\Gamma(d/2-\widetilde\Delta_N)}{\Gamma(\widetilde\Delta_N-1)(\widetilde\Delta_N+N-1)}.
\end{align}
A contraction $I(a,b)=a_{\mu_1}\ldots a_{\mu_N}I^{\mu_1\ldots \mu_N,  \rho_1\ldots \rho_N}
b_{\rho_1}\ldots b_{\rho_N}$ with two vectors $a_\mu$, $b_\mu$  can be written in terms of Gegenbauer polynomials
\begin{align}
I(a,b)=
\frac{N!\,\Gamma(d/2-1)}{2^N\Gamma(N+d/2-1)} (a^2b^2)^{N/2}\, C^{d/2-1}_N \left(\frac{(a\cdot b)}{\sqrt{a^2 b^2}}\right).
\end{align}

The function $\mathcal D_{\widetilde \Delta_N}(x-y)$ can be interpreted as the correlation function of the operators with shadow
scaling dimensions, the shadow operators  ${\widetilde{\mathcal O}}_{\widetilde \Delta_N}^{\vec{\mu}_N}$. If $\Delta_N>d$ the
scaling dimension $\widetilde \Delta_N$ of the shadow operator is negative so that it cannot be realized as a local
operator~\footnote{An example of a shadow operator in a scalar CFT is: $\mathcal O=\varphi$ and $\widetilde O=\delta
S/\delta\varphi$.}. Nevertheless, shadow operators present a convenient technical  tool, see e.g.
\cite{Ferrara:1972uq,Fradkin:1978pp,SimmonsDuffin:2012uy}.

Using \eqref{2point} and \eqref{3point} one can rewrite Eq.~\eqref{first:triangle}
as
\begin{align}
\label{C-relation1}
T^{\vec{\mu}_N}_{\Delta_N}(x_1,x_2,x_3) =
    c''_N C^{\vec{\nu}_N}_{\Delta_N}(x_{12},\partial_x)\mathcal D^{\vec{\nu}_N,\vec{\mu}_N}_{\Delta_N}(x-x_3),
\end{align}
with $c''_N= c_N/c'_N$.
This relation completely determines the coefficient function $C^{\vec{\nu}_N}_{\Delta_N}(x_{12},\partial_x)$,
and can be solved, in principle, order by order in $N$ by expanding both sides in powers of $|x_{12}|$
with fixed $x_{13}^2\sim x_{23}^2$. Using the shadow operator formalism, however, appears to be more efficient.
Let us multiply Eq.~\eqref{C-relation1} by the shadow propagator
$\mathcal D^{\vec{\mu}_N,\vec{\rho}_N}_{\widetilde \Delta_N}(x_3-y)$ and integrate over $x_3$:
\begin{align}
\label{C-relation2}
\int\!d^dx_3\,T^{\vec{\mu}_N}_{\Delta_N}(x_1,x_2,x_3) \mathcal D^{\vec{\mu}_N,\vec{\rho}_N}_{\widetilde \Delta_N}(x_3-y)
&= c''_N C^{\vec{\nu}_N}_{\Delta_N}(x_{12},\partial_x)\int\!d^dx_3\,
\mathcal D^{\vec{\nu}_N,\vec{\mu}_N}_{\Delta_N}(x-x_3)\mathcal D^{\vec{\mu}_N,\vec{\rho}_N}_{\widetilde \Delta_N}(x_3-y)\,.
\end{align}
The integral on the r.h.s. of this relation is given by
Eq.~\eqref{twoDD} whereas the integral on the l.h.s. is fixed (up to a constant) by its transformation properties:
\begin{align}
\int\!d^dx_3\,T^{\vec{\mu}_N}_{\Delta_N}(x_1,x_2,x_3) \mathcal D^{\vec{\mu}_N,\vec{\rho}_N}_{\widetilde \Delta_N}(x_3-y)
=
 r_N(\Delta_N)T_{\widetilde \Delta_N }^{\vec{\rho}_N}(x_1,x_2,y).
\end{align}
The proportionality coefficient  can be calculated explicitly. For completeness we present the result:
\begin{align}\label{rNdelta}
r_N(\Delta_N)& =\pi^{d/2} \,
\frac{\Gamma(d/2-\widetilde \Delta_N)}{\Gamma(\widetilde \Delta_N+N)} \frac{\Gamma(\Delta_N-1+N)}{\Gamma(\Delta_N-1)}
\frac{\Gamma\left(\tilde j_N+\frac12 (\Delta_{1}-\Delta_2)\right)\Gamma\left(\tilde j_N+\frac12 (\Delta_{2}-\Delta_1)\right)}
{\Gamma\left(j_N+\frac12 (\Delta_{1}-\Delta_2)\right)\Gamma\left( j_N+\frac12 (\Delta_{2}-\Delta_1)\right))},
\end{align}
where
$j_N=(\Delta_N+N)/2$, $\tilde j_N=(\widetilde \Delta_N+N)/2$ are conformal spins.
Note that $r_N(\Delta_N) r_N(\widetilde \Delta_N)=D_N(\Delta_N)$, Eq.~\eqref{D_N}.

Finally, taking a Fourier transform of Eq.~\eqref{C-relation2} in $y$, we obtain
\begin{align}\label{C-momentum}
C^{\mu_1\ldots\mu_N}_{\Delta_N}(x_{12}, ip)\,e^{ip\cdot x}=\tilde c_N\int d^dy\, e^{ip\cdot y}
\,T_{\widetilde \Delta_N}^{\mu_1\ldots\mu_N}(x_1,x_2,y)\,,
\end{align}
where $\tilde c_N$ is a product  of the normalization constants defined above
\footnote{
As already mentioned, these constants depend on the theory and the normalization of conformal operators.
For the particular case of QCD they can be related to the coefficient functions in deep-inelastic lepton-hadron
scattering, see a later section.}.

Eq.~\eqref{C-momentum} presents the coefficient function in the product of two scalar operators in a conformal theory including
contributions of all descendants in the form of a Fourier transform (in one variable) of the conformal triangle obtained by the
replacement of the particular primary operator by the corresponding shadow operator. This relation, however, involves the following
subtlety. Remember that the coefficient function is defined as a power expansion, Eq.~\eqref{C-relation1}, at $|x_{12}|\to 0$
keeping $|x_{13}|$ and $|x_{23}|$ fixed. The Fourier integral, Eq~\eqref{C-momentum}, on the other hand, also receives a
contribution from the integration  region of $|y|\sim |x_{12}|$ and for such contributions the summation over the contributions of
descendants and the integration over $y$ cannot be interchanged. The problem can be seen as follows. The coefficient function
$C(x_{12}, \partial)$ is given by the series, roughly speaking, in $(x_{12}^2 \partial^2)$. The derivatives
 act on the function $\mathcal D (x-x_3)$ in Eq.~\eqref{2point} producing factors
${1/(x-x_3)^{2k}}$. The convergence of  the series is therefore controlled by the parameter $|x_{12}/(x-x_3)|$, and it is not
uniform. One can interchange the summation and integration only if $|x_{12}/(x-x_3)| < 1$, i.e. if $|x_{1(2)}|\ll |x_3|$.
Fortunately the integral in Eq.~\eqref{C-momentum} can be calculated in a closed form where the contribution of interest is easily
isolated. The details of this calculation are presented in App.~\ref{App:p-triangle}. Our final result for the OPE of two scalar
operators with the same scaling dimension, $\Delta_1=\Delta_2=\Delta$ reads
\begin{align}\label{OPE=Delta}
\mathcal O_{\Delta}(x_1)\mathcal O_{\Delta}(x_2)&=\sum_{N,\Delta_N}
\frac{c_{N}(\Delta_N)}{|x_{12}|^{2\Delta-t_N}}\sum_{k=0}^N\frac{
N!}{
(N-k)!}  \Gamma(\varkappa_N-k)\left(\frac{x_{12}^{2}}4\right)^k
\notag\\
&\quad
\times \int_0^1du\, (u\bar u)^{j_N-1}
C_k^{\varkappa_N-k}(2u-1)\,
\mathbf I_{\Delta_N+k-d/2}\Big(\sqrt{-u\bar u x_{12}^2\partial^2}\Big)\,
\mathcal{O}_{N,\Delta_N}^{(k)}(x_{21}^u)\,.
\end{align}
Here
\begin{align}
 \bar u = 1-u\,, && x_{12}=x_1-x_2, && x_{21}^u = \bar u x_2 + u x_1\,,
\end{align}
$j_N=(\Delta_N+N)/2$ is the conformal spin of  the operator $\mathcal O_{\Delta_N}^{\mu_1\ldots\mu_N}$,
$\varkappa_N=(d-t_N-1)/2$, where $t_N=\Delta_N-N$ is the operator twist,
$C_k^\lambda(2u-1) $ are Gegenbauer polynomials and $\mathbf I_\nu(x)$
is the modified Bessel function of the first kind (up to normalization)
\begin{align}
\label{Bessel}
\mathbf I_\nu(z)=z^{-\nu} I_{\nu}(z)=2^{-\nu}\sum_{m=0}^\infty \frac{(z^2)^m}{2^{2m}m!\Gamma(\nu+m+1)}
=\frac{2^{-\nu}}{\Gamma(\nu+1)}\,{}_0F_1(\nu+1,z^2/4).
\end{align}
The operator $\mathcal{O}_{N,\Delta_N}^{(k)}$ has twist $t_N+2k$ and is defined as
\begin{align}
\mathcal{O}_{N,\Delta_N}^{(k)}(y)=\partial_y^{\mu_1}\ldots \partial_y^{\mu_k}
{{\mathcal O}_{\Delta_N,\mu_1\ldots\mu_k\mu_{k+1}\ldots\mu_N}(y)}  x_{12}^{\mu_{k+1}}\ldots x_{12}^{\mu_N}.
\end{align}
 The coefficients $c_{N,\Delta_N}$ are not fixed by conformal symmetry and depend on a CFT model.

The operator content in Eq.~\eqref{OPE=Delta} is very explicit:  the summation index $k$
counts applications of the divergence to the primary operator,
and  $m$ in Eq.~\eqref{Bessel} counts applications of the Laplace operator, $\partial^2$, respectively.
This coefficient function was originally derived by Ferrara {\it et al.}~\cite{Ferrara:1971vh}.
The expression presented in~\cite{Ferrara:1971vh} is organized in a different way that
complicates a direct comparison. We have verified that contributions of the operators with $k=0$ agree.

\section{Descendants of leading twist operators in the product of vector currents}\label{sect:vector}

In this Section we generalize the above construction to the OPE of a product of two vector currents. One difference to
the scalar case is that higher-twist conformal operators appearing in the OPE of two vectors are not necessary
symmetric under the interchange of Lorentz indices; to avoid such terms we consider the contribution of
leading twist operators only (and their descendants), which are the only ones necessary for applications to QCD phenomenology.

Scaling dimensions of the leading-twist operators are uniquely determined by their spin $N$,
\begin{align}
\Delta_N=d-2+N+ \gamma_N,
\end{align}
where $\gamma_N$ is the anomalous dimension at the critical point $\gamma_N=\gamma_N(a_s^\star)$.
Hence we can shorten the notation $\mathcal O_{N,\Delta_N}\mapsto\mathcal O_N$.
The twist $t_N$ and conformal spin $j_N$  for the leading twist  operators are given by
\begin{align}
t_N=d-2+\gamma_N, && j_N= \frac12 d -1+N+\frac12 \gamma_N\,.
\end{align}

We are interested in the {OPE} of two conserved vector currents, $j_\mu(x_1) j_\nu(x_2)$. The current conservation, $\partial^\mu
j_\mu=0$, implies that the scaling dimension of $j_\mu$ equals $\Delta=d-1$. Analogously to \eqref{scalar:OPE} we write
\begin{align}
j^\mu(x_1)j^\nu(x_2)=\sum_{N} C^{\mu\nu}_{\mu_1\ldots\mu_N}(x_{12},\partial)\mathcal O_N^{\mu_1\ldots\mu_N}(x)+\ldots,
\end{align}
where $\mathcal O_N^{\mu_1\ldots\mu_N}(x) \equiv \mathcal O_N^{\vec{\mu}_N}(x)$ are leading-twist conformal operators,
$x=(x_1+x_2)/2$ and ellipses stand for the contributions of operators with different quantum numbers. Similar to the scalar case
considered in the previous section the coefficient function $C^{\mu\nu}_{\vec{\mu}_N}(x_{12},\partial)$ can be extracted  from the
three-point correlation function, $\langle j^\mu(x_1)j^\nu(x_2) \mathcal O_N^{\vec{\mu}_N}(x_3)\rangle$. However, in the vector
case the three-point function contains four structures compatible with conformal symmetry { and invariance under the permutation
$(x_1,\mu)\leftrightarrow(x_2,\nu)$
\begin{align}\label{jjO}
\langle j^\mu(x_1)j^\nu(x_2) \mathcal O_N^{\vec{\mu}_N}(x_3)\rangle=
\sum_{n=0}^3 \mathbb{C}_n^N\, \mathbb T_n^{\mu\nu,\vec{\mu}_N}(x_1,x_2,x_3)\,.
\end{align}
The current conservation reduces the number of independent structures to two, so that the answer will contain
two normalization constants that are not determined by symmetry. We will derive explicit expressions for these
constants in terms of the two coefficient functions in deep-inelastic lepton-hadron scattering
that are known to three-loop accuracy (for arbitrary space-dime dimensions)~\cite{Vermaseren:2005qc}.

The structures $\mathbb{T}_n^{\mu\nu,\vec\mu_N} $ in Eq.~\eqref{jjO} can be chosen in a number of ways, there is no commonly
accepted prescription. For our purposes it is convenient to choose the structures that are as closely as possible related to the
scalar triangle \eqref{triangle1} $\Delta_1=\Delta_2=\Delta$ as
\begin{align}\label{T3factor}
T^{\vec{\mu}_N}_{\Delta_N}(x_1,x_2,x_3)= (x_{12}^2)^{-\Delta} X(x_1,x_2,x_3) \,\,\Lambda^{\vec{\mu}_N}(x_1,x_2,x_3)\,,
\end{align}
where
\begin{align}\label{X:def}
 X(x_1,x_2,x_3)=\left(\frac{|x_{12}|}{|x_{13}| |x_{23}|}\right)^{\Delta_N-N}\!\!\!=\
 \left(\frac{x_{12}^2}{x_{13}^2 x_{23}^2}\right)^{t_N/2}.
\end{align}
The factors $X$ and $\Lambda^{\vec{\mu}_N}$ both transform as fields of zero scaling dimension in $x_1,x_2$
under inversion $x_\mu'=x_\mu/x^2$:
\begin{align}
X(x_1,x_2,x_3)&=({{x'}}^2_3)^{t_N} X(x'_1,x'_2,x'_3),
\notag\\
\Lambda^{\mu_1\ldots\mu_N}(x_1,x_2,x_3)&=({{x'}}_3^2)^{N} \eta^{\mu_1}_{\nu_1}(x'_3)\ldots
\eta^{\mu_N}_{\nu_N}(x'_3) \Lambda^{\nu_1\ldots\nu_N}(x'_1,x'_2,x'_3)\, .
\end{align}
Taking into account that a derivative  of an operator with zero scaling dimension,  $\partial^\mu \mathcal O$, transforms as a
primary vector field of dimension $\Delta=1$, we can construct four functions that transform properly in all three coordinates as
follows:
\begin{align}\label{M4}
 \mathbb{T}_0^{\mu\nu,\vec{\mu}_N} &
 =\frac{1}{(x_{12}^2)^{d-1}} \eta^{\mu\nu}(x_{12})\, X(\vec{x}) \, \Lambda^{\vec{\mu}_N}(\vec{x}),
 \notag\\
\mathbb{T}_1^{\mu\nu,\vec{\mu}_N}
  & =\frac{1}{(x_{12}^2)^{d-2}} \partial_1^\mu \partial_2^\nu\,\Big( X(\vec{x}) \,\Lambda^{\vec{\mu}_N}(\vec{x})\Big),
 \notag\\
\mathbb{T}_2^{\mu\nu,\vec{\mu}_N}
 &=\frac{1}{(x_{12}^2)^{d-2}}\eta^{\mu}_{\rho}(x_{12})\eta^{\nu}_{\sigma}(x_{12})
 \partial_1^\sigma \partial_2^\rho\,\Big( X(\vec{x}) \,
 \Lambda^{\vec{\mu}_N}(\vec{x})\Big),
\notag\\
\mathbb{T}_3^{\mu\nu,\vec{\mu}_N} & =\frac{1}{(x_{12}^2)^{d-2}} \Big[\partial_1^\mu\,\Big(  \,
 \Lambda^{\vec{\mu}_N}(\vec{x})\,\partial_2^\nu X(\vec{x})\Big)
 +\partial_2^\nu\,\Big(  \,
 \Lambda^{\vec{\mu}_N}(\vec{x})\,\partial_1^\mu X(\vec{x})\Big)\Big].
\end{align}
Here we used the following notation
\begin{align}
 \partial_k^\mu = \frac{\partial}{\partial x_k^\mu}\,,  && \vec{x} = \{x_1,x_2,x_3\}\,,
\end{align}
and took into account that one only needs structures that are symmetric under the permutation $(x_1,\mu)\leftrightarrow (x_2,\nu)$.
 All these structures can be rewritten in a more conventional form in terms of $\Lambda^\mu$ and
$\eta^{\mu\nu}(x)$ tensors. The advantage of the choice in Eq.~\eqref{M4}, however, is that the first three structures are reduced
essentially to the scalar triangle~\eqref{3point} so that their contribution to the OPE can be obtained without any additional
calculations~\footnote{ The correlation function of vectors currents with non-equal scaling dimensions can be parameterized in a
similar way, applying differential operators in $x_1$ and $x_2$ to the scalar triangle, e.g.,  {$(x_{12}^{-2\Delta_2}
\partial^\nu_2 x_{12}^{2\Delta_2})(x_{12}^{-2\Delta_1}\partial_1^\mu x_{12}^{2\Delta_1}) T^{\vec\mu_N}_{\Delta_N}(x_1,x_2,x_3)$,
 etc.} }.

The four ($N$-dependent) normalization constants $\mathbb{C}_n^N$, $n=0,1,2,3$ in Eq.~\eqref{jjO} are not constrained by conformal
symmetry. However, the current conservation condition $\partial_\mu j^\mu=0$ gives rise to two linear relations { for even $N$,}
\begin{align}
\label{Ward1}
  0&=  2 (\Delta_N\!-\!1) (2 \!-\! d \!+\! \Delta_N\! -\!  N) (\Delta_N\! +\! N)
  \mathbb{C}^N_1 + (\Delta_N\! -\! N) \big[(4 - 3 d) \Delta_N +
    3 \Delta_N^2 - N (d\! +\! N\! -\! 2)\big] \mathbb{C}^N_3 ,
\notag\\[2mm]
  0&=
- (\Delta_N \!+\!N)\mathbb{C}^N_0 + (d\!-\!\Delta_N\!+\!N)(\Delta_N^2-N^2)\mathbb{C}^N_1
- (\Delta_N\!+\!N)\big[\Delta_N^2+ N(N\!-\!2) + d(N\!-\!\Delta_N)\big] \mathbb{C}^N_2
\notag\\&\quad
+  (\Delta_N\!-\!N)\big[2d\Delta_N-2\Delta_N^2 + \Delta_N N + N(N\!+\! 2)\big]\mathbb{C}^N_3\,,
\end{align}
so that only two constants are independent. For odd $N$ all coefficients must vanish identically to ensure current conservation.
One can, for example, use these relations to exclude $\mathbb{C}^N_0$ and
$\mathbb{C}^N_1$ and in this way obtain the three-point function in terms of two conserved structures
\begin{align}
\label{A12}
\langle j^\mu(x_1)j^\nu(x_2) \mathcal O_N^{\vec{\mu}_N}(x_3)\rangle= c_1^N \mathcal A_1^{\mu\nu,\vec{\mu}_N}(x_1,x_2,x_3)+
c_2^N \mathcal A_2^{\mu\nu,\vec{\mu}_N}(x_1,x_2,x_3)\,,
\end{align}
where
$$
\frac{\partial}{\partial x_1^\mu}\mathcal A_k^{\mu\nu,\vec{\mu}_N}(x_1,x_2,x_3)
=\frac{\partial}{\partial x_2^\nu}\mathcal A_k^{\mu\nu,\vec{\mu}_N}(x_1,x_2,x_3)=0\,.
$$
The conserved structures  $A_k^{\mu\nu,\vec{\mu}_N}$ are related to the original ones in Eq.~\eqref{M4} as follows:
\begin{align}
\mathcal A_1^{\mu\nu,\vec{\mu}_N}(\vec{x})&=
a_{10}^N \,\mathbb{T}_0^{\mu\nu,\vec{\mu}_N}(\vec{x})-
 \mathbb{T}_2^{\mu\nu,\vec{\mu}_N}(\vec{x})\, ,
 \notag\\[2mm]
\mathcal{A}_2^{\mu\nu,\vec{\mu}_N}(\vec{x})&= \mathbb{T}_3^{\mu\nu,\vec{\mu}_N}(\vec{x} ) + a_{21}^N\,
\mathbb{T}_1^{\mu\nu,\vec{\mu}_N}(\vec{x}) +
 a_{20}^N\, \mathbb{T}_0^{\mu\nu,\vec{\mu}_N}(\vec{x})\, ,
\end{align}
with the coefficients
\begin{align}\label{ankN}
a_{10}^N &=
 2N(N-1)-(\Delta_N-N)(\widetilde \Delta_N-N),
\notag\\[2mm]
a_{21}^N &=
-\frac{\Delta_N-N}{2(\Delta_N-1)}\Big[3 +
2\frac{N(N+d-2)-\Delta_N}{(\Delta_N+N)(2-N-\widetilde\Delta_N)}\Big],
\notag\\[2mm]
a_{20}^N&=
2(d-2)a_{21}^N+\frac{\Delta_N-N}{2(\Delta_N-1)}
\Big[
4(d-2)+(\widetilde \Delta_N+N)(N+\Delta_N-2)\Big].
\end{align}

As explained in the previous section, the coefficient function in the OPE  can be obtained from the Fourier transform of the
three-point function composed of two vector currents and a spin-$N$ operator with shadow scaling dimension
$\widetilde{\mathcal{O}}_N^{\vec{\mu}_N} \equiv \mathcal{O}_{N,\widetilde{\Delta}_N}^{\vec{\mu}_N}$. We write
\begin{align}\label{jjOtilde1}
\langle j^\mu(x_1)j^\nu(x_2) \widetilde{\mathcal O}_N^{\vec{\mu}_N}(x_3)\rangle=
\sum_{n=0}^3 \widetilde{\mathbb{C}}_n^N\, \widetilde{\mathbb T}_n^{\mu\nu,\vec{\mu}_N}(x_1,x_2,x_3)\,,
\end{align}
where the constants $\widetilde{\mathbb{C}}_n^N$ are related by Ward identities in Eq.~\eqref{Ward1} with the replacement
$\Delta_N \mapsto \widetilde \Delta_N$.
Note that
\begin{align}
\int d^d x_3\,  \mathbb{T}_k^{\mu\nu,\vec{\mu}_N}(x_1,x_2,x_3)\,
\widetilde{\mathcal{D}}^{\vec{\rho}_N}_{\vec{\mu}_N}(x_3-y)= r_N(\Delta_N)
\widetilde{\mathbb T}_k^{\mu\nu,\vec{\rho}_N}(x_1,x_2,y)\,, \qquad k=0,1,2\,,
\end{align}
with the coefficient  $r_N(\Delta_N)$ given in Eq.~\eqref{rNdelta}, but the corresponding integral for $k=3$ is given by a linear
combination of $\widetilde{\mathbb T}_{1}^{\mu\nu,\vec\mu_N}$ and $\widetilde{\mathbb T}_3^{\mu\nu,\vec\mu_N} $. Thus $\widetilde{\mathbb{C}}_n^N \equiv
\mathbb{C}_n^N|_{(\Delta_N\mapsto\widetilde\Delta_N)}  \neq \mathbb{C}_n^N$ whereas in the representation in terms of two conserved
structures in Eq.~\eqref{A12}, we find $\tilde a_{10}^N\equiv a_{10}^N|_{(\Delta_N\mapsto\widetilde\Delta_N)}= a_{10}^N$ but $\tilde a_{2k}^N\equiv a_{2k}^N|_{(\Delta_N\mapsto\widetilde\Delta_N)}\neq
a_{2k}^N$. In this form
\begin{align}\label{jjOtilde}
\langle j^\mu(x_1)j^\nu(x_2) \widetilde{\mathcal O}_N^{\vec{\mu}_N}(x_3)\rangle=
    \tilde c_1^N \widetilde{\mathcal A}^{\mu\nu,\vec{\mu}_N}_{1}(\vec{x})+
    \tilde c_2^N \widetilde{\mathcal A}^{\mu\nu,\vec{\mu}_N}_{2}(\vec{x})\,,
\end{align}
where
\begin{align}\label{coeff:rkN}
\widetilde{\mathbb{C}}_0^N=\tilde c^N_1 \tilde a^N_{10}
+ \tilde c^N_2 \tilde a^N_{20}, && \widetilde{\mathbb{C}}_1^N=\tilde c^N_2 \tilde a^N_{21}, &&
\widetilde{\mathbb{C}}_2^N=-\tilde c^N_1, && \widetilde{\mathbb{C}}_3^N=\tilde c^N_2.
\end{align}

The next step is to calculate the Fourier transform
\begin{align}
 \widetilde{\mathbb{T}}_k^{\mu\nu,\vec{\mu}_N}(x_1,x_2,p) &= \int d^d x_3\, e^{i(p\cdot x_3)}\,
 \widetilde{\mathbb{T}}_k^{\mu\nu,\vec{\mu}_N}(x_1,x_2,x_3)\,,
\end{align}
where one has to select the appropriate contribution corresponding to the integration region $|x_3| \gg |x_{12}|$. One obtains ($N$
is even)
\begin{align}\label{M4p}
 \widetilde{\mathbb{T}}_0^{\mu\nu,\vec{\mu}_N}(x_1,x_2,p) &= \frac{1}{(x_{12}^2)^{d-1}} \eta^{\mu\nu}(x_{12})\,
 \widetilde{\mathbb{S}}^{\vec{\mu}_N}(x_1,x_2,p)\,,
 \notag\\
 \widetilde{\mathbb{T}}_1^{\mu\nu,\vec{\mu}_N}(x_1,x_2,p) &= \frac{1}{(x_{12}^2)^{d-2}} \partial_1^\mu \partial_2^\nu\,
 \widetilde{\mathbb{S}}^{\vec{\mu}_N}(x_1,x_2,p)\,,
 \notag\\
 \widetilde{\mathbb{T}}_2^{\mu\nu,\vec{\mu}_N}(x_1,x_2,p) &=
  \frac{1}{(x_{12}^2)^{d-2}}\eta^{\mu}_{\rho}(x_{12})\eta^{\nu}_{\sigma}(x_{12})
 \partial_1^\sigma \partial_2^\rho\,  \widetilde{\mathbb{S}}^{\vec{\mu}_N}(x_1,x_2,p)\,,
\notag\\
 \widetilde{\mathbb{T}}_3^{\mu\nu,\vec{\mu}_N}(x_1,x_2,p) &=
 \frac{\widetilde \Delta_N-N}{(x_{12}^2)^{d-2}}\biggl\{\Big(\partial^\mu_1 x_{21}^\nu  + \partial^\nu_2 x_{12}^\mu \Big)
\frac1{x_{12}^2} \widetilde{\mathbb{S}}^{\vec{\mu}_N}(x_1,x_2,p)
\notag\\
&\quad
-\partial_2^\nu (x_1^\mu+i\partial_p^\mu)  \widetilde{\mathbb{P}}^{\vec{\mu}_N}(x_1,x_2,p)
-\partial_1^\mu(x_2^\nu+i\partial_p^\nu)  \widetilde{\mathbb{P}}^{\vec{\mu}_N}(x_2,x_1,p)
\biggr\},
\end{align}
where $\partial^\mu_p = \partial/\partial p_\mu$ and
\begin{align}
 \widetilde{\mathbb{S}}^{\vec{\mu}_N}(x_1,x_2,p) &=
 \int d^d x_3\, e^{i(p\cdot x_3)}\, \widetilde X(\vec{x}) \, \Lambda^{\vec{\mu}_N}(\vec{x})\,,
\notag\\
 \widetilde{\mathbb{P}}^{\vec{\mu}_N}(x_1,x_2,p) &=
 \int d^d x_3\, e^{i(p\cdot x_3)}\, \frac{1}{x_{13}^2} \widetilde X(\vec{x}) \, \Lambda^{\vec{\mu}_N}(\vec{x})
\end{align}
are the Fourier transforms of scalar conformal triangles with $\Delta_1=\Delta_2=0$, $\Delta_3=\widetilde{\Delta}_N$ and
$\Delta_1=1$, $\Delta_2=0$, $\Delta_3=\widetilde{\Delta}_N+1$, respectively.
The first function, $\widetilde{\mathbb{S}}^{\vec{\mu}_N}$, has already appeared in the OPE of scalar operators~\eqref{OPE=Delta}.
We obtain
\begin{align}
 \widetilde{\mathbb{S}}^{\vec{\mu}_N}(x_1,x_2,p) &=
 \varpi_N (x_{12}^2)^{\tau_N}
\int_0^1 du\, e^{i(p\cdot x_{21}^u)}  (u\bar u)^{j_N-1}
\sum_{k=0}^N \frac{N!}{(N-k)!} ip^{\mu_1}\ldots ip^{\mu_k}\, x_{12}^{\mu_{k+1}}\ldots x_{12}^{\mu_N}
\notag\\
&\quad
\times \left(\frac{ x_{12}^2}4\right)^k
 \Gamma(\varkappa_N-k)\, C_{k}^{\varkappa_N-k}(u-\bar u)\,
\mathbf I_{\lambda_N+k}\Big( \sqrt{u\bar u p^2 x_{12}^2 }\Big),
\notag\\
 \widetilde{\mathbb{P}}^{\vec{\mu}_N}(x_1,x_2,p) &=
\frac{\varpi_N (x^2_{12})^{\tau_N-1}}{(N \!-\! 2\varkappa_N)(\varkappa_N\! +\! \frac 12)}
\int_0^1\! du\, e^{i(p\cdot x_{21}^u)} u^{j_N-1}\bar u^{j_N-2}
\sum_{k=0}^N \frac{N!}{(N\!-\! k)!} ip^{\mu_1}\!\ldots ip^{\mu_k}\, x_{12}^{\mu_{k+1}}\!\ldots x_{12}^{\mu_N}
\notag\\
&\quad\times
\left(\frac{ x_{12}^2}4\right)^k\Gamma(\varkappa_N\!+\!1\!-\!k)
\Big[ C^{(1+\varkappa_N-k)}_k(u-\bar u) - C^{(1+\varkappa_N-k)}_{k-1}(u-\bar u)\Big]
\notag\\
&\quad\times
\mathbf I_{\lambda_N+k-1}\Big( \sqrt{u\bar u p^2 x_{12}^2} \Big),
\end{align}
where
\begin{align}
\tau_N=t_N/2, && \lambda_N=\Delta_N-d/2, &&\varkappa_N=(1-\gamma_N)/2.
\end{align}
and
\begin{align}
\varpi_N &=\frac1{\sqrt{\pi}} (2\pi)^{d/2} 2^{N-2} \frac{\Gamma(N+\gamma_N)}{\Gamma(1-\gamma_N/2)}
\frac{\sin\pi\gamma_N}{\sin\pi(\gamma_N+d/2-2)}.
\end{align}
In both expressions symmetrization of Lorentz indices and subtraction of traces
is implied.

Collecting all contributions we can write the final result for the
contribution of leading-twist operators and their descendants to
the OPE of two conserved vector currents as
\begin{align}\label{jj-OPE}
j^\mu(x_1) j^\nu(x_2) &= \frac{1}{(x^2_{12})^{d-2}}
\sum_{N,\text{even}}
{2^{\lambda_N}\Gamma(\lambda_N+1)} \int_0^1\!du\, (u\bar u)^{j_N-1}\,\sum_{k=0}^N
\frac{N!\,\Gamma(\varkappa_N-k)}{(N-k)}
{\Gamma(\varkappa_N)}
\notag\\&\quad
\times \Biggl\{ \mathbf D_N^{\mu\nu} (x^2_{12})^{\tau_N}
     \left(\frac{x_{12}^{2}}4\right)^k
C_k^{\varkappa_N-k}(u-\bar u)\, \mathbf I_{\lambda_N+k}\Big(\sqrt{-u\bar u x_{12}^2\partial^2}\Big)
\mathcal{O}_{N}^{(k)}(x_{21}^u)
\notag\\
&\quad {-}\, \widetilde{\mathbb{C}}_3^N
\frac{(\widetilde \Delta_N-N)
 (\varkappa_N-k)}{(N-2\varkappa_N)(\varkappa_N +\frac 12)}
\frac{1}{\bar u}\Big[ C^{(1+\varkappa_N-k)}_k(u-\bar u) - C^{(1+\varkappa_N-k)}_{k-1}(u-\bar u)\Big]
\notag\\
&\quad
\times  \biggl[\partial_2^\nu (x^2_{12})^{\tau_N-1}\left(\frac{x_{12}^{2}}4\right)^k \biggl( x_1^\mu\,
 \mathbf I_{\lambda_N+k-1}\Big(\sqrt{- u\bar u  x_{12}^2\partial^2} \Big)
 \mathcal{O}_{N}^{(k)}(x_{21}^u)
\notag\\
&\quad -\mathbf I_{\lambda_N+k-1}\Big(\sqrt{- u\bar u  x_{12}^2\partial^2} \Big) \mathcal{O}_{N}^{\mu,(k)}(x_{21}^u)\biggr)
+(x_1\leftrightarrow x_2, \mu\leftrightarrow \nu)
\biggr]\Biggr\},
\end{align}
where the differential operator $\mathbf D_N^{\mu\nu}$ is defined as
\begin{align}
\mathbf D_N^{\mu\nu}&= \widetilde{\mathbb{C}}_0^N \frac{\eta^{\mu\nu}(x_{12})}{x_{12}^2} +
\widetilde{\mathbb{C}}_1^N \partial_1^\mu\partial_2^\nu
+ \widetilde{\mathbb{C}}_2^N \eta^{\mu}_{\rho}(x_{12})\eta^{\nu}_{\sigma}(x_{12}) \partial_2^\rho\partial_1^\sigma
+ (\widetilde \Delta_N-N) \widetilde{\mathbb{C}}_3^N
\Big(\partial^\mu_1 x_{21}^\nu  + \partial^\nu_2 x_{12}^\mu \Big)\frac{1}{x_{12}^2}.
\end{align}
The notation $\mathcal{O}_{N}^{(k)}(y)$ and $\mathcal{O}_{N}^{\mu,(k)}(y)$ is used for descendants of the operator
$\mathcal{O}_{N}^{\vec{\mu}_N}(y)$
\begin{align}
\label{eq:O(k)}
\mathcal{O}_{N}^{(k)}(y) & =\partial_y^{\mu_1}\ldots \partial_y^{\mu_k}
\mathcal O_{\mu_1\ldots\mu_k\mu_{k+1}\ldots\mu_N}(y)  x_{12}^{\mu_{k+1}}\ldots x_{12}^{\mu_N},
\notag\\
\mathcal{O}_{N}^{\mu,(k)}(y) & =\partial_y^{\mu_1}\ldots \partial_y^{\mu_k}\, y^\mu\,
\mathcal O_{\mu_1\ldots\mu_k\mu_{k+1}\ldots\mu_N}(y)  x_{12}^{\mu_{k+1}}\ldots x_{12}^{\mu_N}
\end{align}
and the derivatives $\partial^2$ in the expansion of $\mathbf I_\nu$  are understood as
\begin{align}
\partial^2\mathcal{O}_N^{(k)}(x_{21}^u)\equiv
\partial_y^2\mathcal{O}_N^{(k)}(x_{21}^u+y)|_{y=0}.
\end{align}
We also used the following identity
\begin{align}
{i\partial_p^\rho\,e^{ip\cdot y} F(p^2)\Big|_{ip\mapsto\partial_\xi}\,f(\xi)\Big|_{\xi=0}
=-F(-\partial^2) y^\rho\, f(y)\, .}
\end{align}
which holds for arbitrary functions $f$ and $F$.

 The coefficients $ \widetilde{\mathbb{C}}_k^N$, $k=0,1,2,3$ are related by Ward identities \eqref{Ward1} (with the replacement
$\Delta_N\mapsto \widetilde \Delta_N$) which can be solved, e.g., in terms of two independent coefficients as in~\eqref{coeff:rkN}.
Finally, note that the result in Eq.~\eqref{jj-OPE} is written for the OPE in Euclidean space. The transition to Minkowski space
is, however, straightforward.

In Ref.~\cite{Braun:2020yib}  the conformal OPE for conserved vector currents in the leading-twist approximation
is obtained in a different form. We have verified that our expressions agree to the accuracy thereof, see App.~\ref{App:LT}.
%

\subsection{The forward limit}

As well known, the total cross section of deep-inelastic lepton-nucleon scattering (DIS) can be related to the imaginary part of the
Compton amplitude for equal momenta of the initial and the final state hadron --- the forward limit:
\begin{align}
 T_{\mu\nu}(p, q) &= i \int\!d^d x\, e^{iq\cdot x} \langle N(p)| \text{T}\{j_\mu(x) j_\nu(0)\}|N(p)\rangle\, .
\end{align}
The OPE for the Compton tensor is usually written in terms of moments of two structure functions~\cite{Vermaseren:2005qc}
\begin{align}
\label{Compton}
 T_{\mu\nu}(p,q) &=
\sum_{N,\text{even}}\!\!f_N\!\left(\frac{2p\cdot q}{Q^2}\right)^N
\biggl[\left(g_{\mu\nu} - \frac{q_\mu q_\nu}{q^2}\right)C_{L}\left(N,\frac{Q^2}{\mu^2},a_s\right)
\notag\\&\quad
-\left(g_{\mu\nu} - p_\mu p_\nu \frac{4x_B^2}{Q^2} - (p_\mu q_\nu + p_\nu q_\mu)\frac{2x_B}{Q^2}\right)
C_{2}\left(N,\frac{Q^2}{\mu^2},a_s\right)\biggr],
\end{align}
where $Q^2=-q^2$ and  $x_B= Q^2/(2(p\cdot q))$ (we drop electromagnetic charges and the sum over flavors for brevity). The
constants $f_N$ are defined as the reduced matrix elements of leading twist operators
\begin{align}
   \langle p| \mathcal O_N^{\mu_1\ldots\mu_N}(0)|p\rangle = p^{\{\mu_1}\ldots p^{\mu_N\}} f_N \,,
\end{align}
where the normalization of the operators is fixed such that
\begin{align}
   \mathcal O_N^{\mu_1\ldots\mu_N}(0) = i^{N-1}\bar q(0) \gamma^{\{\mu_1} D^{\mu_2}\ldots D^{\mu_N\}} q(0)
   +\,\text{total~derivatives}\,.
\label{eq:normalization}
\end{align}
Here $D^{\mu} = \partial^{\mu} - i g A^{\mu}$ and
$\{\ldots\}$ denotes the symmetrization of all enclosed Lorentz indices and the subtraction
of traces. In this normalization at tree level, we have $C_2(N) =1$ and $C_L(N) =0$.

The coefficient functions $C_{2,L}\left(N,\frac{Q^2}{\mu^2},a_s\right)$ in QCD are usually calculated to fixed order in
perturbation theory and contain logarithms of the scale, terms $\sim (a_s\ln Q^2/\mu^2)^n$. In a conformal theory (in our case QCD
at the critical point) all such logarithms can be resummed and the scale dependence is factorized as
\begin{align}
 C_{2,L}\left(N,\frac{Q^2}{\mu^2},a_s\right) = \left(\frac{\mu}{Q}\right)^{\gamma_N}
  \mathbf{C}_{2,L}\left(N,a_s, \epsilon^\ast\right)\,,
\end{align}
where we added the argument $\epsilon^\ast = (4-d)/2$ to stress that the space-time dimension has to be fine-tuned to the value of
the coupling to ensure vanishing  of the beta-function. The corresponding expressions are known to third order in the strong
coupling~\cite{Vermaseren:2005qc}.

Our result in Eq.~\eqref{jj-OPE} can be mapped to the above representation and in this way all normalization constants fixed by
their relation to DIS. Matrix elements of operators containing total derivatives vanish in the forward limit. Hence to this end we
can neglect contributions of all descendants: in Eq.~\eqref{jj-OPE} only operators $ \mathcal{O}^{(k=0)}_N$ and
$\mathcal{O}^{\mu,(k=0,1)}_N$ have to be kept, the Bessel functions reduce to their values at zero argument, $\mathbf{I}_\nu(0) =
2^{-\nu}/\Gamma(\nu+1)$, and the operators can be moved to the origin, $\mathcal{O}^{(k=0)}_N(x_{12}^u) \mapsto
\mathcal{O}^{(k=0)}_N(0)$.  As a result, the integral over $u$ can easily be taken. Since the Lorentz structure in \eqref{Compton}
is ensured by current conservation, the two coefficients can extracted by considering, e.g., the trace $T_\mu^{\ \mu}$ and the term
$\sim g_{\mu\nu}$ which comes from contributions $\sim g_{\mu\nu}$ and $x_\mu x_\nu$ in  position space. We obtain
\begin{align}\label{trace}
(d-1) \mathbf C_{L}\left(N,a_s,\epsilon^\ast\right)-\left(d - 2\right) \mathbf C_{2}\left(N,a_s,\epsilon^\ast\right) =
\mathbb R_N \Biggl\{\widetilde{\mathbb{C}}_0^N (d-2)
-2( \widetilde{\mathbb{C}}_1^N+ \widetilde{\mathbb{C}}_2^N) t_N \Big(j_N-1 +\frac12 d\Big)\notag\\
+
  \widetilde{\mathbb{C}}_3^N \frac{2(N-1+\frac12\gamma_N)}{(1-\tfrac12\gamma_N)(\gamma_N+ N-1)}
\left[ (N+ 2 d-4 + \gamma_N)\big[\gamma_N N-(1-\gamma_N) t_N\big]{-
N t_N \Big[1 + \frac12 \frac{\gamma_N}{j_N-1}\Big]}\right]
\Biggr\}
\end{align}
and
\begin{align}\label{gmunux}
\mathbf C_{L}\left(N,a_s,\epsilon^\ast\right)- \mathbf C_{2}\left(N,a_s,\epsilon^\ast\right) =
\mathbb{R}_N
\Biggl\{
\biggl[
  \widetilde{\mathbb{C}}_0^N  - t_N  (\widetilde{\mathbb{C}}_1^N+\widetilde{\mathbb{C}}_2^N)
+ \frac{2(2j_N-d)
[- t_N+ \gamma_N \Delta_N ]}{(2-\gamma_N)(N-1+\gamma_N)} \widetilde{\mathbb{C}}_3^N
\biggr]
\notag\\
-\frac{2}{d-\gamma_N}
\biggl[
 \widetilde{\mathbb{C}}_0^N
+ \frac12 t_N (t_N-2)   \widetilde{\mathbb{C}}_1^N
+  2 j_N (j_N-1)  \widetilde{\mathbb{C}}_2^N
- \frac{(t_N-2)(2j_N - d)[- t_N+ \gamma_N\Delta_N ] }{(2-\gamma_N)(N-1+\gamma_N)}
 \widetilde{\mathbb{C}}_3^N
\biggr]
\Biggr\},
\end{align}
where
\begin{align}\label{mathbbRN}
\mathbb{R}_N &=i^N B(j_N,j_N)\, 2^{\gamma_N}\, \pi^{d/2} \frac{\Gamma[\tfrac12\gamma_N+N ]}{\Gamma[\tfrac12 d - \tfrac12\gamma_N]}.
\end{align}
These equations, added by the two relations due to  current conservation \eqref{Ward1}$|_{\Delta_N\mapsto \Delta^\ast_N}$,
define the coefficients $\widetilde{\mathbb{C}}_k^N$ in the conformal expansion \eqref{jj-OPE} in terms of DIS  coefficient
functions at the critical space-time dimension. This system of four linear equations can, of course, be solved explicitly. The
result, however, is rather lengthy  and we present it in the Appendix~\ref{App:LT}.

\subsection{Tree level  result}

The OPE in four-dimensional QCD and the critical QCD in $d=4-2\epsilon$  differs by terms proportional to $\epsilon$ so that
they coincide identically at tree level. The result to this accuracy is much simpler
as compared to the general expression in Eq.~\eqref{jj-OPE}, because one can neglect
all higher-twist terms that are analytic in $x_{12}^2$ and only
produce $\delta$-functions (and derivatives of the  $\delta$-functions) after Fourier transformation to momentum space.
Hence such terms do not contribute to any observables.

Sending $\epsilon \mapsto 0$, $\gamma_N\mapsto 0$  one gets for the coefficients
$\tilde a_{kn}^N \equiv a_{kn}^N|_{\Delta_N\mapsto \widetilde\Delta_N}$ defined in  Eq.~\eqref{ankN}
\begin{align}\label{ankND=4}
\tilde a_{10}^N
        =2(N-1)(N+2),
&&
\tilde a_{20}^N
        =2(N-1)(N+2)/N, &&
\tilde a_{21}^N
        =(N-1)(N+2)/(2N)-2\,.
\end{align}
The tree-level coefficients $\tilde c^N_1$,  $\tilde c^N_2$ in Eq.~\eqref{jjOtilde}
can be found using Eqs.~\eqref{cc-ab1}, \eqref{c1c2(a,b)}
(or, alternatively, \eqref{trace}, \eqref{gmunux}) and the tree-level DIS coefficient functions
$\mathbf{C}_L(N) = 0$, $\mathbf{C}_2(N) = 1$:
\begin{align}
\tilde c_1^N=-\frac{i^N}{2\pi^2}\frac{(2N+1)!}{N!^2 (N+1)!}\,, &&
\tilde c^N_2 =\frac{2N}{(N-1)(N+2)}\, \tilde c^N_1\,.
\end{align}
Thus we get from~\eqref{coeff:rkN}
\begin{align}
&\widetilde{\mathbb C}^{\rm tree}_0=2N(N+1)\tilde c_1^N,&&\widetilde{\mathbb C}_1^{\rm tree}=\frac{N^2-3N-2}{(N-1)(N+2)}\tilde c_1^N,
\notag\\
&\widetilde{\mathbb C}_2^{\rm tree}=-\tilde c_1^N, &&\widetilde{\mathbb C}_3^{\rm tree}=\frac{2N}{(N-1)(N+2)}\tilde c_1^N.
\end{align}
Using these expressions one obtains, after some algebra,
the following expression for the OPE of the time-ordered product of currents in Minkowski space
at tree level
\allowdisplaybreaks{
\begin{align}\label{tree-level-OPE}
\mathrm{T}\{j^{\mu}(x_1)j^\nu(x_2)\}&=
\sum_{N>0,\text{even}} r_N\int_0^1 \!du\, (u\bar u)^{N}\Biggl\{ \frac{1}{(-x_{12}^2+i0)^2}
\Biggl[
(N+1)
g_{\mu\nu}\biggl(1-\frac14 \frac{u\bar u}{N+1} {x_{12}^2}\partial^2\biggr)
\notag\\
&\quad +\frac1{2N} x_{12}^2
\big(\partial_1^\mu\partial_2^\nu-\partial_1^\nu\partial_2^\mu\big)
+\left(1-\frac14  \frac{u\bar u}{N} {x_{12}^2}\partial^2\right)\left(
\frac{\bar u}u x_{21}^\mu\partial_1^\nu+ \frac{u}{\bar u} x_{12}^\nu\partial_2^\mu\right)
\notag\\
&\quad
-\frac14 \frac{u\bar u}{N(N\!+\!1)} {x_{12}^2}\partial^2 \Big(
 x_{21}^\nu\partial_1^\mu\!+\!  x_{12}^\mu\partial_2^\nu\Big)
-\frac{ x_{12}^\mu x_{12}^\nu}{N\!+\!1}  u\bar u\partial^2\biggl(1 - \frac14 \frac{u\bar u}{N\!+\!2} x_{12}^2 \partial^2\biggr)
\Biggr]
\mathcal{O}^{(0)}_N(x_{21}^u)
\notag\\
&\quad
-\frac1{(-x_{12}^2+i0)}\Biggl[-\frac14 N (\bar u -u)\,g_{\mu\nu}
- \frac{\bar u-u}{4(N+1)}
\big( x_{21}^\nu\partial_1^\mu + x_{12}^\mu\partial_2^\nu \big)
\notag\\
&\qquad\qquad
+\frac12\Big(\bar u\,x_{21}^\mu\partial_1^\nu-u\,x_{12}^\nu\partial_2^\mu  \Big)
+\frac{N}{2(N+2)(N-1)}\Big(x_{21}^\nu\partial_1^\mu - x_{12}^\mu\partial_2^\nu \Big)
\notag\\
&\qquad\qquad
+\frac14\frac{N(N^2+N+2)}{(N+1)
(N+2)(N-1)}
\left(\frac{u}{\bar u} x_{12}^\nu\partial_2^\mu
-
\frac{\bar u}u x_{21}^\mu\partial_1^\nu \right)
\notag       \\
&\qquad\qquad
+\frac{x_{12}^\mu x_{12}^\nu}{(-x_{12}^2+i0)} (\bar u-u) \frac{N}{N+1}
\left(1 - \frac12 \frac{u\bar u}{N+2} x_{12}^2 \partial^2 \right)
 \Biggr]\,
 \mathcal{O}^{(1)}_N(x_{21}^u)
 \notag\\
&\quad
 - \frac{x_{12}^\mu x_{12}^\nu}{(-x_{12}^2+i0)}
 \left[\frac{N^2+N+2}{4(N+1)(N+2)} - u\bar u \frac{N(N-1)}{(N+1)(N+2)}
\right] \mathcal{O}_N^{(2)}(x_{21}^u)
\Biggl\}\,+\ldots,
\end{align}
}%
where
\begin{align}
r_N=
-\frac{i^N}{\pi^2} \frac{(2N+1)!}{(N-1)!N!(N+1)!}\,,
\end{align}
and the dots stand for terms regular in $x_{12}^2$. The following identities are useful to collect all such terms:
\begin{align}
C_k^{1/2-k}(1-2u)&
= \frac{\sqrt{\pi} \, 2^{2k-1}}{\Gamma(1/2-k)k!} \Big[ \bar u^k + (-1)^k u^k\Big],
\notag\\
C_k^{3/2-k}(1-2u)&= \frac{\sqrt{\pi}\, 2^{2k-2}}{\Gamma(3/2-k) k!}
\biggl\{ [\bar u^k+(-1)^k u^k] - \frac12 k [\bar u^{k-1}-(-1)^{k-1} u^{k-1}]\biggr\}.
\end{align}
The multiplicatively renormalizable (conformal) leading-twist operators to the required one-loop accuracy take the form
\begin{align}
n_{\mu_1}\ldots n_{\mu_N} \mathcal{O}_N^{\mu_1\ldots\mu_N}(y) =
 \frac{\Gamma(3/2) \Gamma(N)}{\Gamma(N+1/2)} \left(\frac{i\partial_+}{{4}}\right)^{N-1}\bar q(y) \gamma_+
C_{N-1}^{3/2}\left(\tfrac{\stackrel{\rightarrow}{D}_+ - \stackrel{\leftarrow}{D}_+}
                      {\stackrel{\rightarrow}{D}_+ + \stackrel{\leftarrow}{D}_+}\right) q(y)\,,
\end{align}
where we use a shorthand notation $\gamma_+ = \gamma_\mu n^\mu$, etc., with
$n^\mu$ being an auxiliary light-like vector, $n^2=0$,  and the prefactor corresponds to the
normalization in \eqref{eq:normalization}.

Note that the OPE \eqref{tree-level-OPE} contains contributions of the leading-twist operator
$\mathcal{O}_N^{(0)} \equiv \mathcal{O}_N $ (decorated with one or two applications of $\partial^2$),
and also the contributions of the (single and double) divergence of the leading twist
operators $\mathcal{O}_N^{(1)}$ and $\mathcal{O}_N^{(2)}$, cf.~\eqref{eq:O(k)}.
The latter two contributions are nontrivial because matrix elements of  $\mathcal{O}_N^{(1)}$, $\mathcal{O}_N^{(2)}$
over on-shell quarks vanish so that their coefficients cannot be calculated in the standard fashion. The power of
conformal symmetry is that these coefficients are related to the contributions of the leading-twist operators
$\mathcal{O}_N^{(0)}$ by symmetry transformations.

Note also that the OPE \eqref{tree-level-OPE} in this form does not present a complete twist separation in the standard sense.
Expanding the operators around a fixed point, e.g., around $x_2$
\begin{align}
\mathcal{O}_N(x_{21}^u)
& =\sum_m\frac{u^m}{m!} x_{12}^{\nu_1}\ldots x_{12}^{\nu_m} (x_{12})_{\mu_1}\ldots (x_{12})_{\mu_N}
 \partial_{\nu_1}\ldots  \partial_{\nu_m}  \mathcal{O}_N^{\mu_1\ldots\mu_N}(x_2)\, ,
\end{align}
one obtains a series in local operators
$\partial_{\nu_1}\ldots  \partial_{\nu_m}  \mathcal{O}_N^{\mu_1\ldots\mu_N}(x)$ that are traceless in $\{\mu_1\ldots\mu_N\}$
but the traces in  $\{\nu_1\ldots\nu_m\}$ and also mixed trace terms proportional to $\sim g_{\mu_k \nu_l}$ are not subtracted.
This is different to the representation in  Ref.~\cite{Braun:2011zr} where all traces are subtracted and the expansion is
truncated at twist four. We have checked that our result in~\eqref{tree-level-OPE} agrees with the
expression obtained in Ref.~\cite{Braun:2011zr} to this accuracy. Since our result resums all twists,
it is not obvious whether explicit trace subtraction for the contributions involving total derivatives
(which amounts to reshuffling of different contributions without changing the sum)
would be beneficial for applications, so that we do not attempt to do it in this work.

\section{Axial-vector  contributions}\label{sect:axial}

The tree-level contribution of axial-vector operators to the OPE of vector currents can be
obtained starting already from a four-dimensional theory to avoid well-known complications with the
definition of axial operators in non-integer dimensions.
The three-point correlation function of the leading-twist axial conformal operator
$\mathcal O_N^{A,\vec{\mu}_N} \equiv \mathcal O_N^{A,\mu_1\ldots\mu_N}$
with two conserved vector currents is nonzero for odd $N$ only. It
contains only one independent structure and can be written in the following
form (cf.~\cite{Stanev:2012nq}):
\begin{align}
\langle j^\mu(x_1) j^\nu(x_2)\mathcal{O}_N^{A,\vec{\mu}_N}(x_3)\rangle &=\frac{ A_N}{(x_{12}^2)^{d-2}}
 \mathbf K^{\mu\nu}
X(x_1,x_2,x_3)\, \Lambda^{\vec{\mu}}(x_1,x_2,x_3)\,,
\end{align}
where
\begin{align}
 \mathbf K^{\mu\nu} =
 \epsilon^{\mu\alpha\beta\gamma} \eta_{\alpha}^\nu(x_{12}) \eta_{\gamma\gamma'}(x_{12}) \partial_{1,\beta}\partial_2^{\gamma'}
-
\epsilon^{\nu\alpha\beta\gamma} \eta_{\alpha}^\mu(x_{12}) \eta_{\gamma\gamma'}(x_{12}) \partial_{1}^{\gamma'}\partial_{2,\beta}
\end{align}
and $X(x_1,x_2,x_3)$ and $\Lambda^{\vec{\mu}}(x_1,x_2,x_3)$ are defined in
Eqs.~\eqref{X:def} and \eqref{Lambda:def}, respectively.

The two terms in $\mathbf K^{\mu\nu}$ are in fact equal to each other.
This is easy to see taking into account that~\cite{Fradkin:1978pp}
\begin{align}\label{etaepsilon}
{\eta^{\mu}}_{\mu'}(x)\ldots {\eta^{\sigma}}_{\sigma'}(x)\epsilon^{\mu'\nu'\rho'\sigma'}=-\epsilon^{\mu\nu\rho\sigma},
\end{align}
leading to
\begin{align}
\epsilon^{\mu\alpha\beta\gamma} \eta_{\alpha}^\nu(x_{12}) \eta_{\gamma\gamma'}(x_{12}) \partial_{1,\beta}\partial_2^{\gamma'}
=-\epsilon^{\nu\alpha\beta\gamma} \eta_{\alpha}^\mu(x_{12}) \eta_{\gamma\gamma'}(x_{12}) \partial_{1}^{\gamma'}\partial_{2,\beta}.
\end{align}
We prefer to write $\mathbf K^{\mu\nu}$ in the above form to make the symmetry under $(x_1,\mu)\leftrightarrow(x_2,\nu)$ explicit.

The calculation is similar to the vector case. Repeating the same steps  we obtain
the contribution of axial-vector operators to the OPE of two vector currents at tree level
\begin{align}
\text{T}\{j_\mu(x_1) j_\nu(x_2)\}_A &= \frac1{ (-x_{12}^2+i0)^2}
\sum_{N,\text{odd}} \widetilde{\mathbb C}_A^N  \frac{2^{N+2} N!}{\sqrt \pi} \int_0^1du\, (u\bar u)^{N}\,\sum_{k=0}^N\frac{
N!}{
(N-k)!} \Gamma\left(\frac12-k\right)
\notag\\
&\quad
\times  \mathbf K_{\mu\nu}\,\left(\frac{x_{12}^{2}}4\right)^{k+1} C_k^{\frac12-k}(u-\bar u)\,
\mathbf I_{N+k}\Big({-u\bar u x_{12}^2\partial^2}\Big)
\mathcal{O}_{N}^{A,(k)}(x_{21}^u).
\end{align}
 The coefficients $\widetilde{\mathbb C}_A^N$ are not constrained by the symmetries and have to be
fixed from, e.g., the forward limit.

Assuming that axial-vector operators are normalized as
\begin{align}
   \mathcal O_N^{A,\mu_1\ldots\mu_N}(0) = i^{N-1}\bar q(0)\gamma^{\{\mu_1}D^{\mu_2}\ldots D^{\mu_N\}}\gamma_5 q(0)
   +\,\text{total~derivatives}\,,
\label{eq:normalization-axial}
\end{align}
and dropping all terms that produce $\delta$-functions in momenta after the Fourier transform, one obtains
\begin{align}
\label{OPE-axial-tree}
\text{T}\{j_\mu(x_1) j_\nu(x_2)\}_A &= \frac1{ (-x_{12}^2+i0)^2}\sum_{N,\text{odd}} r^A_N
\int_0^1 du\, (u\bar u)^{N}
\Biggl\{\epsilon_{\mu\nu\beta\gamma} x_{12}^\beta
\Biggl[ \biggl[N\left(\frac{u}{\bar u}\partial_{2}^\gamma -\frac{\bar u}{u}\partial_{1}^\gamma\right)
\notag\\
&\quad
-\frac14 \frac{ u\bar u x_{12}^2\partial^2}{(N+1)}
\Big( \partial_{2}^\gamma-\partial_{1}^\gamma + (N+1)\Big(\frac{u}{\bar u}\partial_{2}^\gamma
-\frac{\bar u}{u}\partial_{1}^\gamma
\Big)\Big)
\biggr]\mathcal O_{N}^{A,(0)}(x_{21}^u)
\notag\\
&\quad
-\frac14 \frac{N\, x_{12}^2}{N+1}
\biggl[
\Big(N (\bar u-u)+\frac{\bar u}u\Big)\frac{u}{\bar u}\partial_{2}^\gamma
+
\Big(N(u-\bar u)+\frac{u}{\bar u}\Big)\frac{\bar u}{u}\partial_{1}^\gamma
\biggr]
\mathcal O_{N}^{A,(1)}(x_{21}^u)\Biggr]
\notag\\
&\quad
-\Big (x_{12,\nu}\epsilon_{\mu\alpha\beta\gamma}
+x_{12,\mu} \epsilon_{\nu\alpha\beta\gamma}
\Big)  x_{12}^\alpha
\biggl[\left(1-\frac 1 4\frac{u\bar u x_{12}^2\partial^2}{N+1}
\right)
\partial_{1}^\beta \partial_{2}^\gamma\mathcal O_{N}^{A,(0)}(x_{21}^u)
\notag\\
&\quad
{-}\frac14 \frac {N (\bar u-u)}{N+1}  x_{12}^2
\partial_{1}^\beta \partial_{2}^\gamma\mathcal O_{N}^{A,(1)}(x_{21}^u)\biggr]
\Biggr\}+\ldots
\end{align}
with
\begin{align}
r^A_N=\frac{i^{N-1}}{2\pi^2}\frac{(2N+1)!}{N!^2 (N+1)!}\, .
\end{align}
We have checked that this expression agrees with the result in
Ref.~\cite{Braun:2011zr} to twist-four accuracy (i.e. the difference is twist-five and higher).

\section{Summary}\label{sect:summary}

We have derived an all-order expression for the coefficient functions of the leading-twist (vector) operators and their
descendants, obtained by adding total derivatives, in the expansion of a product of two electromagnetic currents,
Eq.~\eqref{jj-OPE}, in conformal QCD at the Wilson-Fisher critical point in non-integer space-time dimensions. This expression can
be viewed as a generalization of the classical result by  Ferrara {\it et al.}~\cite{Ferrara:1971vh} who considered the OPE of the
product of { two scalar currents} in a conformal theory. Our expression for the scalar case, Eq.~\eqref{OPE12}, is more explicit
and seems to be  more useful for applications.

The motivation for this study is, most importantly, the application to deeply-virtual Compton scattering (DVCS), which is the
gold-plated processes for spatial imaging of partons inside the nucleon. Descendants of the leading twist operators give rise to
corrections to DVCS amplitudes that are proportional to powers of the nucleon mass {$m$} and invariant momentum transfer $t$ to the
target. As the spatial position of partons is Fourier conjugate to the momentum transfer, the resolving power of DVCS is bound by
the maximum $t$ that can be used in the analysis.  Since factorization for DVCS { beyond the leading-twist accuracy requires the
inclusion of} power corrections in $\sqrt{-t}/Q$ and {$m/Q$}, theoretical control over these corrections is of paramount
importance. They can significantly impact the extraction of GPDs from the data and have to be taken into
account~\cite{Defurne:2015kxq,Defurne:2017paw}.

The rationale for using methods based on conformal symmetry, as already discussed in the Introduction, is that
matrix elements of some of the relevant operators vanish for free quarks so that their coefficients cannot be calculated
in a standard manner. This task was already addressed in Refs.~\cite{Braun:2011zr,Braun:2011dg,Braun:2012bg,Braun:2012hq}
using a different technique. The necessary operator-level expressions were derived in~\cite{Braun:2011zr,Braun:2011dg}
to twist-four accuracy. A new contribution of this paper is that our result includes all twists.
Such all-order results are especially important for the newly emerging
subject of coherent DVCS from light nuclei \cite{Hattawy:2017woc}, in which case one needs to prove that QCD factorization is not
spoiled by the nucleus mass corrections, terms $\sim m_A/Q$.

At the tree level, the OPE in physical  QCD in $d=4$ coincides identically with the OPE in conformal QCD in $d=4-2\epsilon$ so
that our operator-level results to this accuracy, Eqs.~\eqref{tree-level-OPE} and \eqref{OPE-axial-tree} can be taken directly
to calculate the DVCS amplitudes, cf.~\cite{Braun:2014sta}.
This work is in progress. Beyond the tree level, conformal QCD expressions, in general, have to be modified by terms
proportional to the QCD beta-function. Such corrections enter only at $\mathcal{O}(a_s^2)$ (NNLO) for leading twist,
whereas for higher twists there can be an additional subtlety related to different twist definition
in $d=4$ and $d=4-2\epsilon$ dimensions. In this case a separate study is needed.

\section*{Acknowledgments}
\addcontentsline{toc}{section}{Acknowledgments}

We are grateful to Sergey Derkachov for collaboration in the early stages of the project and to Slava Rychkov for a helpful
discussion. This work was supported by Deutsche Forschungsgemeinschaft (DFG) through the Research Unit FOR 2926, ``Next Generation
pQCD for Hadron Structure: Preparing for the EIC'', project number 40824754,
 DFG grant MO 1801/1-3, and RSF project No 19-11-00131. Y.J. also acknowledges the support of DFG, grants
BR 2021/7-2 and SFB TRR 257.


\appendix
\addcontentsline{toc}{section}{Appendices}
\renewcommand{\theequation}{\Alph{section}.\arabic{equation}}
\renewcommand{\thesection}{{\Alph{section}}}
\renewcommand{\thetable}{\Alph{table}}
\setcounter{section}{0} \setcounter{table}{0}

\section*{Appendices}

\section{Notations}\label{App:notations}

For readers' convenience we collect here a list of notations used throughout this work.

We consider conformal QCD in $d=4-2\epsilon$ space-time dimensions at the Wilson-Fischer critical point;
$a_s= \alpha_s/(4\pi)$ is the strong coupling and the criticality condition is $\epsilon = -\beta_0 a_s + \mathcal{O}(a_s^2)$
with $\beta_0=11/3 N_c-2/3 n_f$.
We use the following notations:
\begin{align}
 &\bar u = 1-u\,, \qquad \qquad x_{ij}= x_i-x_j\,, \qquad  x_{ij}^u = \bar u x_i + u x_j\,,
\end{align}
\begin{align}
&\eta^{\mu\nu}(x) = g^{\mu\nu}- \frac{2 x^\mu x^\nu}{x^2}\,,
\qquad
\Lambda^\mu(x_1,x_2,x_3) =\frac{x_{13}^\mu}{x_{13}^2}-\frac{x_{23}^\mu}{x_{23}^2}\,.
\end{align}
We use vector notation for a set of Lorentz indices, e.g.,
\begin{align}
   \vec{\mu}_N = \{\mu_1,\mu_2,\ldots, \mu_N\}\,.
\end{align}
Symmetrization over all indices in the set and subtraction of traces is implied. $\Delta_N$ stands for the scaling dimension of a
local (conformal) operator $\mathcal{O}_N^{\vec{\mu}_N}$  with Lorentz spin $N$. For leading twist operators in QCD
\begin{align}
  \Delta_N = N + d-2+ \gamma_N\,,
\end{align}
where $\gamma_N$ is the anomalous dimension, $j_N$ stands for the conformal spin and $t_N$ for twist:
\begin{align}
 j_N &= \frac12 (\Delta_N+N) = \frac12 d -1 +N + \frac12 \gamma_N\,,
\notag\\
 t_N &= \Delta_N -N = d-2+ \gamma_N\,.
\end{align}
We also use the following shorthand notations:
\begin{align}
\tau_N=\frac12 t_N , && \lambda_N=\Delta_N-\frac12 d , && \varkappa_N= \frac12 (d-t_N-1) =  \frac12 (1-\gamma_N)\,.
\end{align}
The quantities with a ``tilde'' generally refer to expressions calculated with the replacement of the
scaling dimension $\Delta_N$ by the shadow dimension
\begin{align}
 \widetilde \Delta_N = d- \Delta_N\,.
\end{align}
The (partial) derivatives with a subscript $1,2$  stand for
\begin{align}
   \partial_{k,\mu} = \frac{\partial}{\partial x_k^\mu}\,,\qquad k=1,2\,,
\end{align}
whereas the derivatives without a subscript are understood as derivatives over the position  of the operator in the
OPE, e.g.,
\begin{align}
\partial^2\mathcal{O}_N^{(k)}(x_{21}^u)\equiv
\partial_y^2\mathcal{O}_N^{(k)}(x_{21}^u+y)|_{y=0}.
\end{align}

\section{Fourier transform of the conformal triangle}\label{App:p-triangle}
In this Appendix we calculate the integral
\begin{align}
\int d^dy\, e^{ip\cdot y} \,T_{\Delta_N}^{\mu_1\ldots\mu_N}(x_1,x_2,y)\,,
\end{align}
where $T_{\Delta_N}^{\mu_1\ldots\mu_N}$ is the conformal triangle Eq.~\eqref{triangle1}. It is convenient to contract all vector
indices with an auxiliary light-like vector $n^\mu$, $n^2=0$, so  that the symmetrization over all indices and trace subtraction is
ensured:
\begin{align}
  T^{(n)}_{\Delta_N}(x_1,x_2,y)  &= n_{\mu_1}\ldots n_{\mu_N} T_{\Delta_N}^{\mu_1\ldots\mu_N}(x_1,x_2,y)
\notag\\&=
\frac{\Lambda^N_n(x_1,x_2,y)}{|x_{12}|^{\Delta_1+\Delta_2-\Delta_N+N}
|x_1-y|^{\Delta_1+\Delta_N-N-\Delta_2} |x_2-y|^{\Delta_2+\Delta_N-N-\Delta_1}},
\end{align}
with $\Lambda_n=n\cdot\Lambda(x_1,x_2,y)$.  The following representation proves to be useful:
\begin{align}\label{disk-rep}
  T^{(n)}_{\Delta_N}(x_1,x_2,y) &=
\frac{2^{-N}}{|x_{12}|^{\Delta_1+\Delta_2-\Delta_N+N}}
\int_{\mathfrak{D}} \!D_{j_1}z_1 \int_{\mathfrak{D}}\! D_{j_2}z_2\,\frac{\bar z_{12}^N}{|x_1-y-z_1n|^{4j_1}|x_2-y-z_2n|^{4j_2}},
\end{align}
where $\bar z_{12} = \bar z_1-\bar z_2$ { is the complex conjugate of $z_{12}$} and
\begin{align}
\label{smallj12}
  4j_1 = \Delta_1+\Delta_N-N-\Delta_2\,,  &&  4j_2 = \Delta_2+\Delta_N-N-\Delta_1\,.
\end{align}
The integration over $z_1$, $z_2$ goes over the unit disk $|z_k|\leq 1$ in the complex plane
\begin{align}
  \mathfrak{D} = \{ z\in \mathbb{C}, |z|<1\}
\end{align}
and the integration measure is defined as
\begin{align}
D_j z=\frac{2j-1}{\pi} (1-|z|^2)^{2j-2} d^2z\,.
\end{align}
To verify~\eqref{disk-rep} it is sufficient to note that for any function $f(z)$ analytic inside the unit disk (see e.g.
\cite{MR1770752}) the following identity holds
\begin{align}
f(w)=\int_{\mathfrak{D}} \! D_j z\, (1-w \bar z)^{-2j} f(z)\,.
\end{align}
Thus we have,
\begin{align}
   \int_{\mathfrak{D}} D^{j_1}z_1 \frac{\bar z_{12}^N}{(x_1-y -z_1n)^{4 j_1}} &=
\int_{\mathfrak{D}} D_{j_1}z_1 \frac{\bar z_{12}^N}{[(x_1-y)^2 - 2 z_1(n\cdot(x_1-y)))]^{2j_1}}
\notag\\&=
\frac{1}{[(x_1-y)^2]^{2j_1}} \int_{\mathfrak{D}} D_{j_1}z_1 \frac{\bar z_{12}^N}{[1- 2 z_1(n\cdot(x_1-y))/ (x_1-y)^2]^{2j_1}}
\notag\\ &=
\frac{1}{[(x_1-y)^2]^{2j_1}}  \left( 2 \frac{n\cdot(x_1-y)}{(x_1-y)^2} - \bar z_2\right)^N
\end{align}
and repeating the same trick for the integral over $z_2$ one arrives at \eqref{disk-rep}.

Next, we combine the two $y$-dependent factors in \eqref{disk-rep} using Feynman's formula
\begin{align}
T^{(n)}_{\Delta_N}(x_1,x_2,p)&=\int d^dy\, e^{ipy}\, T^{(n)}_{\Delta_N}(x_1,x_2,y)
\notag\\&=
\frac{1}{2^N B(2j_1,2j_2)}\frac1{|x_{12}|^{\Delta_1+\Delta_2 -\Delta_N+N}}\int_0^1du\, u^{2j_1-1}\bar u^{2j_2-1}e^{i(p\cdot x_{21}^u)}
\notag\\
&\quad
\times
\int_{\mathfrak{D}} D_{j_1} z_1 D_{j_2} z_2 \,\bar z_{12}^N  e^{-iz_{21}^u (p\cdot n)}
 \int d^dy\frac{e^{ipy}}{[y^2+u\bar u x_{12}^2-2 u\bar u z_{12}(x_{12}\cdot n)]^{2(j_1+j_2)}}\, ,
\end{align}
where $B(2j_1,2j_2)$ is Euler's beta function.
The integrals over unit discs can now be taken by expanding the denominator in a power series in $z_{12}$
and using
\begin{multline}
\int_{\mathfrak{D}} D_{j_1} z_1 D_{j_2} z_2 \,\bar z_{12}^N\,  z_{12}^k\, e^{-iz_{21}^u(p\cdot n)}
=
(p\cdot n)^{-k} \left(i\frac{d}{du}\right)^k \int_{\mathfrak{D}} D_{j_1} z_1 D_{j_2} z_2 \,\bar z_{12}^N\,   e^{-iz_{21}^u (p\cdot n)}
\\
=[-i(p\cdot n)]^{N-k}  N!\frac{\Gamma(2j_1)\Gamma(2j_2)}{\Gamma(N+2j_1)\Gamma(N+2j_2)}\frac{d^k}{du^k}
 P^{(2j_2-1,2j_1-1)}_N\left(2u-1\right),
\end{multline}
where $P^{(a,b)}_N(x)$ are Jacobi's polynomials. Collecting all powers, after some algebra one gets
\begin{align}
T^{(n)}_{\Delta_N}(x_1,x_2,p)&=\frac{N!}{\Gamma(N+2j_1)\Gamma(N+2j_2)}\sum_{k=0}^N\frac{\Gamma(k+2j_1+2j_2)
\Gamma(N+k+2j_1+2j_2-1)}{k!\Gamma(N+2j_1+2j_2-1)}
\notag\\
&\quad \times  \left(\frac12\right)^{N-k}\frac{(x_{12}\cdot n)^k [-i (p\cdot n) ]^{N-k}}{|x_{12}|^{\Delta_1+\Delta_2-\Delta_N+N}}
\int_0^1du\, u^{2j_1-1+k}\bar u^{2j_2-1+k}\, e^{i(p\cdot  x_{21}^u)}
\notag\\&\quad
\times P^{(2j_2-1+k,2j_1-1+k)}_{N-k}\left(2u\!-\!1\right)
\int d^dy \frac{e^{ip\cdot y}}{(y^2+u\bar u x_{12}^2)^{2j_1+2j_2+k}}.
\end{align}
The remaining Fourier integral $\int d^dy$ can be taken in terms of the modified Bessel function of the second kind
\begin{align}\label{Mcdonald}
\int d^dy \frac{e^{ip\cdot y}}{(y^2+u\bar u x_{12}^2)^{\alpha}}=
\frac{(2\pi)^{d/2}}{\Gamma(\alpha)} 2^{1-\alpha} (u\bar u x_{12}^2)^{d/2-\alpha}
\left(\sqrt{{u\bar u x_{12}^2}{p^2}}\right)^{\alpha-d/2}
K_{d/2-\alpha}\left(\sqrt{u\bar u x_{12}^2p^2}\right).
\end{align}
Thus we obtain
\begin{align}
T^{(n)}_{\Delta_N}(x_1,x_2,p)&=
\frac{(2\pi)^{d/2} N!}{\Gamma(N+2j_1)\Gamma(N+2j_2)}
\frac1{|x_{12}|^{\Delta_1+\Delta_2 + \Delta_N -d +N}}\sum_{k=0}^N\frac{
\Gamma(N+k+2j_1+2j_2-1)}{k!\Gamma(N+2j_1+2j_2-1)}
\notag\\&
\times  2^{1-2j_1-2j_2-N}(x_{12}\cdot n)^k  \left[- i (p\cdot n) x_{12}^2\right]^{N-k}
\int_0^1\!du\, u^{d/2-2j_2-1}\bar u^{d/2 -2j_1-1}\, e^{i(p\cdot x_{21}^u)}
\notag\\
&\times
P^{(2j_2-1+k,2j_1-1+k)}_{N-k}\left(2u\!-\!1\right) \left(\sqrt{{u\bar u x_{12}^2}{p^2}}\right)^{2j_1+2j_2+k-d/2}
\!\!\!\!\!\!\!\! K_{d/2-2j_1-2j_2-k}\left(\sqrt{u\bar u x_{12}^2p^2}\right).
\end{align}
Replacing in this expression $\Delta_N\mapsto \widetilde \Delta_N$, changing the
summation index $k\mapsto N-k$ and using \eqref{smallj12}, one gets for the shadow triangle
\begin{align}
\label{shadow-answer}
T^{(n)}_{\widetilde\Delta_N}(x_1,x_2,p)&=
\frac{\mathcal{N}}{|x_{12}|^{\Delta_1+\Delta_2 - t_N}}\sum_{k=0}^N
\frac{\Gamma(2\varkappa_N-k)}{(N-k)!}
(x_{12}\cdot n)^{N-k} [-
i(p\cdot n) x_{12}^2]^{k}
\notag\\
&\quad \times
\int_0^1du\, (u\bar u)^{N-1+\frac12 t_N } \left(\frac u{\bar u}\right)^{\Delta_{12}}
 P^{(\varkappa_N- \Delta_{12}-k-\frac12,\varkappa_N+ \Delta_{12}-k-\frac12)}_{k}\left(2u-1\right)
\notag\\
&\quad \times \left(\sqrt{{u\bar u x_{12}^2}{p^2}}\right)^{d/2-\Delta_N-k}
K_{\Delta_N+k-d/2}\left(\sqrt{u\bar u x_{12}^2p^2}\right)\,e^{i(p\cdot x_{21}^u)},
\end{align}
where $\Delta_{12} = \frac12 (\Delta_1-\Delta_2)$, $t_N=\Delta_N-N$, $\varkappa_N=\frac12(d-t_N-1)$ and
\begin{align}
\mathcal{N} =\frac{(2\pi)^{d/2} N! }{2^{\widetilde \Delta_N-1}\Gamma(\widetilde \Delta_N-1)
\Gamma(\varkappa_N- \Delta_{12}+\frac12)\Gamma(\varkappa_N + \Delta_{12}+\frac12)
}.
\end{align}

Coming back to the integral in \eqref{Mcdonald}, the result can be written in the form of the following
asymptotic expansion
\begin{align}
&\left(\sqrt{u\bar u x_{12}^2 p^2}\right)^{-\nu}  K_\nu\left(\sqrt{u\bar u x_{12}^2 p^2}\right)=
\nonumber\\ =&\
2^{\nu-1}\biggl\{\frac{\Gamma[-\nu]}{4^\nu}
\sum\limits_{n=0}^\infty \frac{\Gamma[1+\nu]\,4^{-n}}{n! \Gamma[1+\nu+n]} (u\bar u x_{12}^2 p^2)^n
+
\frac{\Gamma[\nu]}{(u\bar u x_{12}^2 p^2)^{\nu}}
 \sum\limits_{n=0}^\infty \frac{\Gamma[1-\nu] 4^{-n}}{n! \Gamma[1-\nu+n]} (u\bar u x_{12}^2 p^2)^n\biggr\},
\end{align}
where the first and the second series correspond to the expansion of the integrand in the
regions of $|y|\gg |x_{12}|$ and  $|y|\ll |x_{12}|$, respectively.
This decomposition corresponds to using the following representation:
\begin{align}\label{eq:BesselK}
K_\nu(z)
= -\frac{\pi}{2\sin\pi\nu}\Big[ \, I_\nu(z)- \, I_{-\nu}( z)\Big]\, ,
\end{align}
where the first term $\propto I_\nu(z)$ sums up contributions of large $y$. Only this term has to be kept in the construction
of the coefficient function of the OPE.
Thus we replace in \eqref{shadow-answer}
\begin{align}
 z^{-\nu} K_\nu(z) \mapsto \mathbf I_\nu(z) = z^{-\nu} I_{\nu}(z)=2^{-\nu}\sum_{m=0}^\infty \frac{(z^2)^m}{2^{2m}m!\Gamma(\nu+m+1)}
=2^{-\nu}\,{}_0F_1(\nu+1,z^2/4)
\end{align}
(the prefactor in Eq.~\eqref{eq:BesselK} $-\pi/(2\sin\pi(\Delta_N+k-d/2))\mapsto (-1)^k\times{\cal N}$ with ${\cal N}$ being an
overall normalization which can be dropped) and end up with the following expression for the OPE of two scalar currents,
Eq.~\eqref{scalar:OPE}:
\begin{align}\label{OPE12}
\mathcal O_{1}(x_1)\mathcal O_{2}(x_2)&=\sum_{N,\Delta_N} \frac{c_{N,\Delta_N}}{|x_{12}|^{\Delta_1+\Delta_2-t_N}}\sum_{k=0}^N\frac{
\Gamma(2\varkappa_N-k)}{
(N-k)!}  |x_{12}|^{2k}
\int_0^1du\, (u\bar u)^{N-1+\frac12 t_N} \left(\frac u{\bar u}\right)^{\Delta_{12}}
\notag\\
&\quad\times
 P^{(\varkappa_N- \Delta_{12}-k-\frac12,\varkappa_N+\Delta_{12}-k-\frac12)}_{k}\left(2u\!-\!1\right)
\mathbf I_{\Delta_N+k-d/2}\Big(\sqrt{-u\bar u x_{12}^2\partial^2}\Big)\,\mathcal{O}_N^{(k)}(x_{21}^u),
\end{align}
where
\begin{align}
\mathcal{O}_N^{(k)}(y)=\partial_y^{\mu_1}\ldots \partial_y^{\mu_k}
\mathcal O_{\mu_1\ldots\mu_N}(y)  x_{12}^{\mu_{k+1}}\ldots x_{12}^{\mu_N}
\end{align}
and the derivatives $\partial^2$ in the expansion of $\mathbf I_\nu$  are understood as
\begin{align}
\partial^2\mathcal{O}_N^{(k)}(x_{21}^u)\equiv
\partial_y^2\mathcal{O}_N^{(k)}(x_{21}^u+y)|_{y=0}.
\end{align}
The result quoted in Eq.~\eqref{OPE=Delta} in the main text corresponds to the particular case $\Delta_1=\Delta_2=\Delta$, hence
$\Delta_{12}=0$.

Since the expansion~\eqref{OPE12} has to be invariant under permutation $(x_1,\Delta_1)\leftrightarrow (x_2,\Delta_2)$,
 only the operators with even spin $N$ contribute to the sum.
Note that in the case  $\Delta_1=\Delta_2=\Delta$ the dependence on the scaling dimension of the currents, $\Delta$,
only comes through the normalization constant $c_{N,\Delta_N}$ and the scaling factor $1/x_{12}^{2\Delta}$.

In the special case $\varkappa_N=1/2$ which corresponds to the scaling dimension
$\Delta_N=d+N-2$, the $\Gamma$-function with the argument $2\varkappa_N-k$ becomes singular for $k>0$.
In a realistic CFT, the only existing operator with such property is the energy-momentum tensor, $T_{\mu\nu}$, ($N=2$).
Its divergence $\partial^\mu T_{\mu\nu}=0$ and also $\partial^\mu \partial^\nu T_{\mu\nu}=0$ (terms with $k=1,2$)
have to be dropped. A more detailed discussion of this case can be found in Refs.~\cite{Ferrara:1971vh,Ferrara:1973yt}.

\section{Leading twist approximation}\label{App:LT}

In~Ref.~\cite{Braun:2020yib} the conformal OPE for two conserved vector
currents in the leading twist approximation (that is neglecting all higher-twist
descendants) is written in the form:
\begin{align}\label{COPE}
\text{T}\,\big\{j^\mu(x_1) j^\nu(x_2)\big\} & =
\sum_{N,\text{even}}\frac{\mu^{\gamma_N}}{(-x_{12}^{2}+i0)^{\frac{d}{2}-\frac12\gamma_N}}
 \int_0^1 du
 \Biggl\{ -\frac12 A_N(u) \eta^{\mu\nu}(x_{12})
  + B_N(u) g^{\mu\nu}
 \notag\\
&\quad
 +C_N(u) x_{12}^\nu\partial_1^\mu - C_N(\bar u) x_{12}^\mu
 \partial_2^\nu
+D_N(u)x_{12}^2 \partial_1^\mu\partial_2^\nu \Biggr\}
\mathcal O_N^{(0)}(x_{21}^u)\,,
\end{align}
where $\mu$ is the scale parameter.
The functions $A_N(u)$,  $B_N(u)$, $C_N(u)$  and $D_N(u)$
are given by the following expressions
\begin{align}
A_N(u) &= a_N\, u^{j_N-1}\bar u^{j_N-1}\,, \qquad\qquad B_N(u) = b_N\, u^{j_N-1} \bar u^{j_N-1}\, ,
\notag\\
C_N(u) & = u^{N-1}\int_u^1\frac{dv}{v^N}v^{j_N} \bar v^{j_N-2} \left(c_N -\frac{b_N}v\right),
 \notag\\
D_N(u)  &= -\frac1{N\!-\!1} \int_0^1\!{dv}( v  \bar v)^{j_N-1}\left[\theta(v-u) \left(\frac{u} v\right)^{N-1}
 + \theta(\bar v-\bar u) \left(\frac{\bar u}{\bar v}\right)^{N-1}  \right]\left(d_N -\frac{c_N-b_N}{2 v \bar v} \right),
\label{ABCD}
\end{align}
with the coefficients $a_N, b_N, c_N$, $d_N$ that obey the following linear equations:
\begin{align}
\left(j_N-1\right)\, a_N & = (d-\gamma_N)\left(c_N -b_N\right)\, ,
\notag\\
(d-2-\gamma_N)\, d_N &= -\frac12 a_N (N-j_N)-\gamma_N b_N + \big(j_N-2+ d-\gamma_N\big) (c_N -b_N)\, .
\end{align}
Comparing \eqref{jj-OPE} and \eqref{COPE}, we get for $a_N$ and $b_N$:
\begin{align}\label{ab-cc}
\mu^{\gamma_N}a_N & = 2(\tau_N+2\varkappa_N)\left[2\tau_N\, \tilde c_1^N +
\Big(N-1+\frac12 \gamma_N\Big)
    \frac{\tau_N^3-(\tau_N-\gamma_N)[j_N(j_N-1)-\tau_N]}{(N-2\varkappa_N)(j_N-1)(\varkappa_N+1/2)}\tilde c_2^N\right],
\notag\\
\mu^{\gamma_N} b_N & = 2\big[ j_N(j_N-1)+\tau_N\big]\, \tilde c_1^N
- {\tau_N } \Big(N-1+\frac12 \gamma_N\Big) \frac{
(\tau_N-2\gamma_N) j_N(j_N-1)-\tau_N^2(\tau_N+2\varkappa_N)
}{(N-2\varkappa_N)(j_N-1)(\varkappa_N+1/2)} \tilde c_2^N ,
\end{align}
where $\tilde c_{1,2}^N$ are defined in Eqs.~\eqref{A12}, \eqref{coeff:rkN}. These linear relations can easily be solved for
\begin{align}
\label{c1c2(a,b)}
\tilde c_2^N &=-\frac{(j_N-1)(N-1+\gamma_N)(1-\frac12\gamma_N)}{2N(N-1)(\tau_N-\gamma_N)(\tau_N+1-\gamma_N)(N-1+\frac12\gamma_N)
(j_N+\tau_N)(j_N-1+\tau_N)}
\notag\\ &\quad \times
    \mu^{\gamma_N}\biggl[a_N(j_N(j_N-1)+\tau_N)-2  b_N \tau_N (\tau_N+1-\gamma_N)\biggr],
\notag\\
\tilde c_1^N& =\frac1{2\tau_N}\biggl[\frac12\frac{\mu^{\gamma_N} a_N}{\tau_N+1-\gamma_N}
-\Big(N-1+\frac12 \gamma_N\Big)
    \frac{\tau_N^3-(\tau_N-\gamma_N)[j_N(j_N-1)-\tau_N]}{(N-1+\gamma_N)(j_N-1)(1-\frac12\gamma_N)}\tilde c_2^N\biggr].
\end{align}
and, finally, the coefficients $a_N$ and $b_N$ can be related to the coefficient functions in DIS~\cite{Braun:2020yib}
\begin{align}\label{cc-ab1}
\mu^{\gamma_N} \mathbb R_N {a_N}
&=\mathbf C_{L}\left(N,a_s,\epsilon_\ast\right)\frac{\tau_N+1-\gamma_N}{N+\frac12\gamma_N}
\,,
\notag\\
\mu^{\gamma_N}\mathbb R_N b_N &=
-\mathbf C_{1}\left(N,a_s,\epsilon_\ast\right)
    + \frac12\mathbf C_{L}\left(N,a_s,\epsilon_\ast\right)
\frac{\tau_N-\gamma_N}{N+\frac12\gamma_N}.
\end{align}
where $\mathbf C_{1}=\mathbf C_{2}-\mathbf C_{L}$, and $\mathbb R_N$ is defined in Eq.~\eqref{mathbbRN}.
Substituting these expressions in \eqref{c1c2(a,b)} and using Eqs.~\eqref{coeff:rkN} one obtains the four coefficient functions
$\widetilde{\mathbb{C}}_k^N$ in the OPE \eqref{jj-OPE} in terms of two DIS coefficient functions.
Alternatively, the same expressions can be obtained by solving Eqs.~\eqref{trace}, \eqref{gmunux} and the
Ward identities \eqref{Ward1}$|_{\Delta_N\mapsto \Delta^\ast_N}$.

We have checked that the expressions for $C _N(u)$ and $D_N(u)$ in Eq.~\eqref{COPE} that follow from~\eqref{jj-OPE} agree with Eq.~\eqref{ABCD}.


\bibliography{BDJM}

\providecommand{\href}[2]{#2}\begingroup\raggedright\begin{thebibliography}{10}

\bibitem{Dudek:2012vr}
J.~Dudek et~al., \emph{{Physics Opportunities with the 12 GeV Upgrade at
  Jefferson Lab}},
  \href{http://dx.doi.org/10.1140/epja/i2012-12187-1}{\emph{Eur. Phys. J. A}
  {\bfseries 48} (2012) 187},
  [\href{https://arxiv.org/abs/1208.1244}{{\ttfamily 1208.1244}}].

\bibitem{Kou:2018nap}
{\scshape Belle-II} collaboration, W.~Altmannshofer et~al., \emph{{The Belle II
  Physics Book}}, \href{http://dx.doi.org/10.1093/ptep/ptz106}{\emph{PTEP}
  {\bfseries 2019} (2019) 123C01},
  [\href{https://arxiv.org/abs/1808.10567}{{\ttfamily 1808.10567}}].

\bibitem{Accardi:2012qut}
A.~Accardi et~al., \emph{{Electron Ion Collider: The Next QCD Frontier}:
  {Understanding the glue that binds us all}},
  \href{http://dx.doi.org/10.1140/epja/i2016-16268-9}{\emph{Eur. Phys. J. A}
  {\bfseries 52} (2016) 268},
  [\href{https://arxiv.org/abs/1212.1701}{{\ttfamily 1212.1701}}].

\bibitem{Braun:2014sta}
V.~M. Braun, A.~N. Manashov, D.~M{\"u}ller and B.~M. Pirnay, \emph{{Deeply
  Virtual Compton Scattering to the twist-four accuracy: Impact of finite-$t$
  and target mass corrections}},
  \href{http://dx.doi.org/10.1103/PhysRevD.89.074022}{\emph{Phys. Rev. D}
  {\bfseries 89} (2014) 074022},
  [\href{https://arxiv.org/abs/1401.7621}{{\ttfamily 1401.7621}}].

\bibitem{Braun:2011zr}
V.~Braun and A.~Manashov, \emph{{Kinematic power corrections in off-forward
  hard reactions}},
  \href{http://dx.doi.org/10.1103/PhysRevLett.107.202001}{\emph{Phys. Rev.
  Lett.} {\bfseries 107} (2011) 202001},
  [\href{https://arxiv.org/abs/1108.2394}{{\ttfamily 1108.2394}}].

\bibitem{Braun:2011dg}
V.~Braun and A.~Manashov, \emph{{Operator product expansion in QCD in
  off-forward kinematics: Separation of kinematic and dynamical
  contributions}}, \href{http://dx.doi.org/10.1007/JHEP01(2012)085}{\emph{JHEP}
  {\bfseries 01} (2012) 085},
  [\href{https://arxiv.org/abs/1111.6765}{{\ttfamily 1111.6765}}].

\bibitem{Braun:2012bg}
V.~Braun, A.~Manashov and B.~Pirnay, \emph{{Finite-t and target mass
  corrections to DVCS on a scalar target}},
  \href{http://dx.doi.org/10.1103/PhysRevD.86.014003}{\emph{Phys. Rev. D}
  {\bfseries 86} (2012) 014003},
  [\href{https://arxiv.org/abs/1205.3332}{{\ttfamily 1205.3332}}].

\bibitem{Braun:2012hq}
V.~Braun, A.~Manashov and B.~Pirnay, \emph{{Finite-t and target mass
  corrections to deeply virtual Compton scattering}},
  \href{http://dx.doi.org/10.1103/PhysRevLett.109.242001}{\emph{Phys. Rev.
  Lett.} {\bfseries 109} (2012) 242001},
  [\href{https://arxiv.org/abs/1209.2559}{{\ttfamily 1209.2559}}].

\bibitem{Defurne:2015kxq}
{\scshape Jefferson Lab Hall A} collaboration, M.~Defurne et~al.,
  \emph{{E00-110 experiment at Jefferson Lab Hall A: Deeply virtual Compton
  scattering off the proton at 6 GeV}},
  \href{http://dx.doi.org/10.1103/PhysRevC.92.055202}{\emph{Phys. Rev. C}
  {\bfseries 92} (2015) 055202},
  [\href{https://arxiv.org/abs/1504.05453}{{\ttfamily 1504.05453}}].

\bibitem{Defurne:2017paw}
M.~Defurne et~al., \emph{{A glimpse of gluons through deeply virtual compton
  scattering on the proton}},
  \href{http://dx.doi.org/10.1038/s41467-017-01819-3}{\emph{Nature Commun.}
  {\bfseries 8} (2017) 1408},
  [\href{https://arxiv.org/abs/1703.09442}{{\ttfamily 1703.09442}}].

\bibitem{Hattawy:2017woc}
{\scshape CLAS} collaboration, M.~Hattawy et~al., \emph{{First Exclusive
  Measurement of Deeply Virtual Compton Scattering off $^4$He: Toward the 3D
  Tomography of Nuclei}},
  \href{http://dx.doi.org/10.1103/PhysRevLett.119.202004}{\emph{Phys. Rev.
  Lett.} {\bfseries 119} (2017) 202004},
  [\href{https://arxiv.org/abs/1707.03361}{{\ttfamily 1707.03361}}].

\bibitem{Polyakov:2018zvc}
M.~V. Polyakov and P.~Schweitzer, \emph{{Forces inside hadrons: pressure,
  surface tension, mechanical radius, and all that}},
  \href{http://dx.doi.org/10.1142/S0217751X18300259}{\emph{Int. J. Mod. Phys.
  A} {\bfseries 33} (2018) 1830025},
  [\href{https://arxiv.org/abs/1805.06596}{{\ttfamily 1805.06596}}].

\bibitem{Tanaka:2018wea}
K.~Tanaka, \emph{{Operator relations for gravitational form factors of a spin-0
  hadron}}, \href{http://dx.doi.org/10.1103/PhysRevD.98.034009}{\emph{Phys.
  Rev. D} {\bfseries 98} (2018) 034009},
  [\href{https://arxiv.org/abs/1806.10591}{{\ttfamily 1806.10591}}].

\bibitem{Ferrara:1971vh}
S.~Ferrara, A.~Grillo and R.~Gatto, \emph{{Manifestly conformal covariant
  operator-product expansion}},
  \href{http://dx.doi.org/10.1007/BF02770435}{\emph{Lett. Nuovo Cim.}
  {\bfseries 2S2} (1971) 1363--1369}.

\bibitem{Ferrara:1971zy}
S.~Ferrara, R.~Gatto and A.~Grillo, \emph{{Conformal invariance on the light
  cone and canonical dimensions}},
  \href{http://dx.doi.org/10.1016/0550-3213(71)90333-6}{\emph{Nucl. Phys. B}
  {\bfseries 34} (1971) 349--366}.

\bibitem{Ferrara:1973yt}
S.~Ferrara, A.~Grillo and R.~Gatto, \emph{{Tensor representations of conformal
  algebra and conformally covariant operator product expansion}},
  \href{http://dx.doi.org/10.1016/0003-4916(73)90446-6}{\emph{Annals Phys.}
  {\bfseries 76} (1973) 161--188}.

\bibitem{Wilson:1973jj}
K.~Wilson and J.~B. Kogut, \emph{{The Renormalization group and the epsilon
  expansion}},
  \href{http://dx.doi.org/10.1016/0370-1573(74)90023-4}{\emph{Phys. Rept.}
  {\bfseries 12} (1974) 75--199}.

\bibitem{Braun:2018mxm}
V.~Braun, A.~Manashov, S.~Moch and M.~Strohmaier, \emph{{Conformal symmetry of
  QCD in $d$-dimensions}},
  \href{http://dx.doi.org/10.1016/j.physletb.2019.04.027}{\emph{Phys. Lett. B}
  {\bfseries 793} (2019) 78--84},
  [\href{https://arxiv.org/abs/1810.04993}{{\ttfamily 1810.04993}}].

\bibitem{Braun:2020yib}
V.~Braun, A.~Manashov, S.~Moch and J.~Schoenleber, \emph{{Two-loop coefficient
  function for DVCS: Vector contributions}},
  \href{https://arxiv.org/abs/2007.06348}{{\ttfamily 2007.06348}}.

\bibitem{Ferrara:1972uq}
S.~Ferrara, A.~Grillo, G.~Parisi and R.~Gatto, \emph{{The shadow operator
  formalism for conformal algebra. Vacuum expectation values and operator
  products}}, \href{http://dx.doi.org/10.1007/BF02907130}{\emph{Lett. Nuovo
  Cim.} {\bfseries 4S2} (1972) 115--120}.

\bibitem{Poland:2018epd}
D.~Poland, S.~Rychkov and A.~Vichi, \emph{{The Conformal Bootstrap: Theory,
  Numerical Techniques, and Applications}},
  \href{http://dx.doi.org/10.1103/RevModPhys.91.015002}{\emph{Rev. Mod. Phys.}
  {\bfseries 91} (2019) 015002},
  [\href{https://arxiv.org/abs/1805.04405}{{\ttfamily 1805.04405}}].

\bibitem{Costa:2011mg}
M.~S. Costa, J.~Penedones, D.~Poland and S.~Rychkov, \emph{{Spinning Conformal
  Correlators}}, \href{http://dx.doi.org/10.1007/JHEP11(2011)071}{\emph{JHEP}
  {\bfseries 11} (2011) 071},
  [\href{https://arxiv.org/abs/1107.3554}{{\ttfamily 1107.3554}}].

\bibitem{Polyakov:1970xd}
A.~M. Polyakov, \emph{{Conformal symmetry of critical fluctuations}},
  {\emph{JETP Lett.} {\bfseries 12} (1970) 381--383}.

\bibitem{Fradkin:1978pp}
E.~Fradkin and M.~Palchik, \emph{{Recent Developments in Conformal Invariant
  Quantum Field Theory}},
  \href{http://dx.doi.org/10.1016/0370-1573(78)90172-2}{\emph{Phys. Rept.}
  {\bfseries 44} (1978) 249--349}.

\bibitem{SimmonsDuffin:2012uy}
D.~Simmons-Duffin, \emph{{Projectors, Shadows, and Conformal Blocks}},
  \href{http://dx.doi.org/10.1007/JHEP04(2014)146}{\emph{JHEP} {\bfseries 04}
  (2014) 146}, [\href{https://arxiv.org/abs/1204.3894}{{\ttfamily 1204.3894}}].

\bibitem{Vermaseren:2005qc}
J.~Vermaseren, A.~Vogt and S.~Moch, \emph{{The Third-order QCD corrections to
  deep-inelastic scattering by photon exchange}},
  \href{http://dx.doi.org/10.1016/j.nuclphysb.2005.06.020}{\emph{Nucl. Phys. B}
  {\bfseries 724} (2005) 3--182},
  [\href{https://arxiv.org/abs/hep-ph/0504242}{{\ttfamily hep-ph/0504242}}].

\bibitem{Stanev:2012nq}
Y.~S. Stanev, \emph{{Correlation Functions of Conserved Currents in Four
  Dimensional Conformal Field Theory}},
  \href{http://dx.doi.org/10.1016/j.nuclphysb.2012.07.027}{\emph{Nucl. Phys. B}
  {\bfseries 865} (2012) 200--215},
  [\href{https://arxiv.org/abs/1206.5639}{{\ttfamily 1206.5639}}].

\bibitem{MR1770752}
B.~C. Hall, \emph{Holomorphic methods in analysis and mathematical physics},
  in \emph{First {S}ummer {S}chool in {A}nalysis and {M}athematical {P}hysics
  ({C}uernavaca {M}orelos, 1998)}, vol.~260 of \emph{Contemp. Math.},
  pp.~1--59.
\newblock Amer. Math. Soc., Providence, RI, 2000.
\newblock \href{http://dx.doi.org/10.1090/conm/260/04156}{DOI}.

\end{thebibliography}\endgroup

\bibliographystyle{JHEP}


\end{document}